\documentclass[preprint2]{aastex}

\usepackage{natbib}
\usepackage{url}
\usepackage[section]{placeins}
\usepackage{graphics,graphicx}
\usepackage[caption=false]{subfig}
\usepackage{pdflscape}
\usepackage{amsmath}

\newcommand{\feii}{[Fe \textsc{ii}]}

\newcommand{\brg}{Br$\gamma$}
\newcommand{\brd}{Br$\delta$}
\newcommand{\molhy}{H$_2$}
\newcommand{\paa}{Pa$\alpha$}

\def\ltsima{$\; \buildrel < \over \sim \;$}
\def\simlt{\lower.5ex\hbox{\ltsima}}
\def\gtsima{$\; \buildrel > \over \sim \;$}
\def\simgt{\lower.5ex\hbox{\gtsima}}

\shorttitle{Black Hole Growth in Mergers}
\shortauthors{Medling et al.}

\begin{document}


\title{Following Black Hole Scaling Relations Through Gas-Rich Mergers} 


\author{Anne M. Medling\altaffilmark{1}, 
Vivian U\altaffilmark{2}, 
Claire E. Max\altaffilmark{3},  
David B. Sanders\altaffilmark{4},
Lee Armus\altaffilmark{5},  
Bradford Holden\altaffilmark{3},
Etsuko Mieda\altaffilmark{6},
Shelley A. Wright\altaffilmark{6,7},
James E. Larkin\altaffilmark{8}
}

\altaffiltext{1}{Research School of Astronomy \& Astrophysics, Mount Stromlo Observatory, Australia National University, Cotter Road, Weston, ACT 2611, Australia; anne.medling@anu.edu.au}

\altaffiltext{2}{Department of Physics and
   Astronomy, University of California, Riverside, 900 University
   Avenue, Riverside, CA 92521, USA}
   
\altaffiltext{3}{Department of Astronomy \& Astrophysics, University of California, Santa Cruz, 1156 High Street, Santa Cruz, CA 95064, USA}

\altaffiltext{4}{Institute for Astronomy, University of Hawaii,
   2680 Woodlawn Drive, Honolulu, HI 96822, USA}

\altaffiltext{5}{Spitzer Science Center, California Institute of
    Technology, 1200 E. California Blvd., Pasadena, CA 91125, USA}

\altaffiltext{6}{Department of Astronomy \& Astrophysics, University of Toronto, 50 St. George Street. Toronto, ON M5S 3H4, Canada
}

\altaffiltext{7}{Dunlap Institute for Astronomy \& Astrophysics, 50 St. George Street. Toronto, ON M5S 3H4, Canada
}

\altaffiltext{8}{Department of Physics and Astronomy, University of California, Los Angeles, CA 90095-1547, USA}
   



\begin{abstract}

We present black hole mass measurements from kinematic modeling of high-spatial resolution integral field spectroscopy of the inner regions of 9 nearby (ultra-)luminous infrared galaxies in a variety of merger stages.  These observations were taken with OSIRIS and laser guide star adaptive optics on the Keck I and Keck II telescopes, and reveal gas and stellar kinematics inside the spheres of influence of these supermassive black holes.  We find that this sample of black holes are overmassive ($\sim10^{7-9}$ M$_{\sun}$) compared to the expected values based on black hole scaling relations, and suggest that the major epoch of black hole growth occurs in early stages of a merger, as opposed to during a final episode of quasar-mode feedback.  The black hole masses presented are the dynamical masses enclosed in $\sim$25pc, and could include gas which is gravitationally bound to the black hole but has not yet lost sufficient angular momentum to be accreted.  If present, this gas could in principle eventually fuel AGN feedback or be itself blown out from the system.

\end{abstract}




\section{Introduction}

Virtually every massive galaxy hosts a supermassive black hole in its core.  These central black holes exhibit tight correlations with properties of their host galaxies' bulges \citep{KormendyGebhardt01}: bulge mass \citep{KormendyRichstone95,Magorrian98}, bulge luminosity \citep{MarconiHunt03}, and bulge stellar velocity dispersion \citep{Tremaine02,FerrareseMerritt00,Gebhardt00}.  
Recently these correlations were re-explored using the wealth of new black hole mass measurements ($10^{6-10} M_{\sun}$) and galaxy parameters by \citet{McConnell13} and \citet{GrahamScott13}.  However, the discovery that the $M_{BH} - \sigma_{*}$ relation may evolve with redshift \citep[e.g.][]{Zhang12} has led to the suggestion that the most basic scaling relation may instead be with total stellar mass \citep{Jahnke09,Cisternas11}; in this case, the evolution of the $M_{BH} - \sigma_{*}$ relation with redshift indicates the changing fractions of mass in galaxy bulges (which contribute to $\sigma_{*}$) versus disks (which do not).  See \citet{KormendyHo13} for a detailed review of our current understanding of black hole scaling relations.

The mechanism through which black hole masses correlate with galaxy properties has been canonically associated with gas-rich galaxy mergers \citep[e.g.][]{Hopkins06}.  Gravitational torques funnel the gas into their centers, triggering two phenomena: an intense burst of star formation to feed the bulge, and accretion of gas on to the black holes in the centers of each galaxy.  It has been postulated that black hole growth can regulate this process through AGN feedback \citep{Springel05,Hopkins10} via massive winds that evacuate the gas from the galaxy on short timescales, cutting off star formation and future black hole growth.  This sense of self-regulation has been confirmed observationally by \citet{KauffmannHeckman09}, who find that the Eddington ratio of a sample of AGNs depends on the supply of cold gas in the galaxy.  If there is plenty of cold gas, the accretion rate does not depend on the quantity of gas available; if the supply of cold gas is limited, the accretion rate depends on the rate at which stellar winds provide fuel for the AGN.  Though the detailed mechanisms causing these correlations are still unconfirmed, star formation and black hole growth are fed by the same reservoir of inflowing gas; their growth histories are intertwined.  It is likely that these two processes compete for fuel in a predictable fashion.  

\begin{figure}[ht]
\includegraphics[scale=.7, trim=1cm 0cm 0cm 0cm]{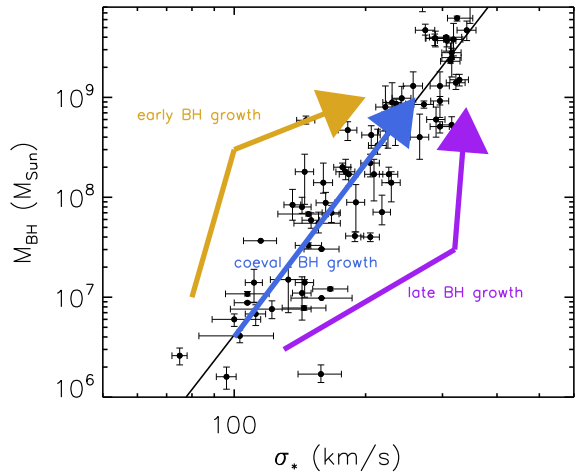}
\caption{$M_{BH} - \sigma_{*}$ relation for isolated galaxies from \cite{McConnell13} (black) with three possible evolutionary tracks for merging galaxies overlaid.  If the black hole grows first or more quickly than the galaxy bulge, mergers would lie above the relation (as shown by the gold arrows). If the black hole growth lags the bulge growth and is responsible for curtailing evolution (e.g. quenching through AGN feedback), mergers would lie below the relation (as shown by the purple arrows). If instead the black hole and the bulge grow in lockstep, the mergers would remain on the relation (as shown by the blue arrow).
}
\label{scalingrelationtracks}
\end{figure}

To understand this interplay, it is critical to look at systems in the midst of this increased fueling.  One set of such galaxies are gas-rich mergers, which tend to have extreme bursts of star formation and a higher incidence of AGN activity \citep[e.g.][]{Sanders88,Sanders96,Veilleux02,IshidaPhD,Ellison13,Koss13}.  During such a merger, does the black hole grow first, leaving the stars to slowly consume the remaining gas? Or is star formation quenched once the black hole reaches a bright quasar phase of extreme growth? The position of a merger on black hole scaling relations would indicate the relative growth timescales, and confirm whether the putative quasar-mode feedback occurs at the end of a merger (see Figure~\ref{scalingrelationtracks} for the schematic example for the $M_{BH} - \sigma_{*}$ relation.).  As star formation happens on few-kiloparsec spatial scales, accreting gas must lose less angular momentum to fuel a starburst than to feed a black hole.  Subsequent models therefore suggest that a black hole would grow substantially only after star formation has quenched itself and the galaxy bulge is in place \citep{Cen12}, or at least that the peak black hole accretion time occurs later in a merger relative to peak star formation \citep{Hopkins12}.  This scenario would predict that gas-rich mergers would fall below black hole scaling relations.

Merger-driven galaxy evolution is not a complete explanation for all black hole growth, however.  Though some AGN studies find a correlation with major mergers \citep{Koss10, Ellison11}, AGNs found with other selection techniques and at different redshift ranges do not show a higher rate of merging than field galaxies \citep{Cisternas11_merger,Kocevski12}.  Instead of mergers bringing in gas, some systems probably are undergoing secular evolution, accreting their gas directly from the cold intergalactic medium.  Bar and spiral disk instabilities are capable of dissipating sufficient angular momentum from this gas to fuel black hole growth \citep{Hopkins10_BHaccretion,Hopkins11}.  The dividing line between these two processes is not yet well understood; here we focus on understanding the black hole growth due to major mergers.

Nearby (Ultra-)Luminous InfraRed Galaxies \citep[(U)LIRGs;][]{Sanders96} are an excellent sample with which to study such gas-rich mergers.  These galaxies have infrared luminosities upwards of $10^{11} L_{\sun}$ ($10^{12} L_{\sun}$ for ULIRGs), generally caused by a starburst and/or an AGN heating up dust.  Their infrared luminosities correlate with merger rate, star formation rate, and AGN fraction \citep{Ellison13}.  In fact, in the local universe, such strong infrared activity is almost exclusively triggered by major mergers: \citet{Veilleux02} showed that in a complete sample of IRAS ULIRGs, 117 out of 118 galaxies are in the midst of strong tidal interactions.  

In a companion paper \citep{nucleardisks}, we have studied the kinematics of both gas and stars in the inner kiloparsecs of a sample of (U)LIRGs, finding that nuclear disks on scales of a few tens to hundreds of parsecs are common.  In two other papers \citep{medling11, mrk273}, we demonstrated a technique that uses high spatial resolution integral field spectroscopy to measure black hole masses.  With kinematic maps that resolve inside the sphere of influence of a black hole, the complex and unrelaxed large-scale dynamics are less important, and black hole masses can be measured to within a factor of a few.  We note that this technique measures the unresolved central mass, which includes both the black hole and its accretion disk, and in some cases may also include a reservoir of gas feeding the accretion disk.  In \cite{mrk273}, the black hole mass in Mrk273 N measured with this technique was consistent with the measurement made by OH maser kinematics \citep{Klockner04}.

Using a robust technique such as this to measure black hole masses in these gas-rich mergers is important because more traditional methods of black hole mass measurements rely on assumptions that aren't valid in the case of galaxy mergers.  Three-integral orbital superposition models \citep[as in][]{Gult_nuker,Siopis} are able to use large-field kinematics to separate different components to the mass profile of the galaxy; however, this approach requires a dynamically relaxed system and is used therefore in isolated galaxies.  Another successful black hole mass measurement technique is reverberation mapping \citep[e.g.][and references therein]{Reverb}, which measures the time lag between flux variations of the continuum and the lines in the broad line region.  Since the cores of (U)LIRGs are so dusty, the broad line regions are too obscured to view.

Obtaining data at spatial resolutions sufficiently high to resolve inside the sphere of influence of a supermassive black hole at the typical redshifts of local (U)LIRGs ($z \lesssim 0.1$) requires adaptive optics systems, which are becoming available at an increasing number of ground-based observatories.  Though the Hubble Space Telescope has excellent resolution in the visible bands, its relatively small mirror size limits the resolution at the longer wavelengths ($>2 \mu$m) necessary to look through the dust in these galactic nuclei; even at the longest wavelength available to the Wide Field Camera 3 ($H$-band), these nuclei are still sometimes obscured.  In order to achieve high spatial resolution in $K$-band, large ground-based telescopes have employed adaptive optics systems which measure turbulence in the Earth's atmosphere and use a deformable mirror to correct for the resulting distortions.  These distortion measurements require references, either a natural guide star (NGS AO) or a laser guide star plus a fainter natural ``tip-tilt'' star (LGS AO).  The addition of laser guide star adaptive optics has increased the area of the sky observable with this technique.  

Throughout this paper we have adopted a cosmology of $H_0 =
70$\,km\,s$^{-1}$\,Mpc$^{-1}$, $\Omega_{\rm m}$ = 0.28, and
$\Omega_\Lambda$ = 0.72 \citep{Hinshaw09}.  In \S\ref{obs} we present our data and reduction techniques.  In \S\ref{kinematicfitting} we briefly describe the kinematic fitting techniques demonstrated in \citet{medling11} and \citet{mrk273}.  In \S\ref{results} we present the black hole masses measured from several tracers and in \S\ref{scalingrelations} compare them to black hole scaling relations.  \S\ref{conclusions} contains our conclusions.

\section{Observations}
\label{obs}

\subsection{The Sample}
We have selected 9 gas-rich merging galaxies in which to measure the black hole masses using gas and stellar kinematics.  These galaxies represent the subset of merging galaxies presented in \citet{nucleardisks} for which high quality kinematics data exist.  The parent sample was drawn from the Great Observatories All-Sky LIRG Survey \citep[GOALS;][]{Armus09}, which targeted about two hundred of the brightest infrared galaxies in the sky (log(L$_{IR}$/L$_{\sun}$) $>$ 11.0).  Our targets were selected from that sample for the following criteria: available archival $B$- and $I$-band imaging from the Hubble Space Telescope Advanced Camera for Surveys; visible double or extended cores or clumps in the central five arcseconds (within our field of view); large-scale morphology consistent with that of a major merger; observable from Keck Observatory and with an appropriate guide star ($\leq$ 18th magnitude in $R$, $\leq 60$\arcsec~separation).

Our sample comprises (U)LIRGs involved in major gas-rich galaxy mergers, with a bias towards later stages of merging.  These galaxies are nearby, with redshifts $\leq 0.05$, which is required to achieve spatial resolutions of $\lesssim30$ pc per pixel ($\lesssim50$ pc per resolution element).  

\subsection{The Data}
We obtained near-infrared integral field spectroscopy of the central kiloparsec of 9 merging galaxies with OSIRIS, the OH-Suppressing InfraRed Imaging Spectrograph \citep{Larkin06} on the W.~M. Keck I (after August 2012) and II (before August 2012) 10-meter telescopes.  All data presented here uses the 0.035 arcsec spaxel$^{-1}$ plate scale.  This high spatial resolution is enabled by the Keck Observatory LGS AO system \citep{Wiz00,vanDam04,Wiz06,vanDam06}.  

Our OSIRIS observations were typically comprised of observing sets of object-sky-object; each exposure was ten minutes (5 minutes for Keck I observations).  Observations were taken either in the broad-band $K$ filter (Kbb: 1.965-2.381 $\mu$m) or in a narrow band targeting specific lines (Hn4: 1.652-1.737 $\mu$m; Kn5: 2.292-2.408 $\mu$m).  When possible, the primary target lines were the CO (2-0) and (3-1) bandheads at 2.293 $\mu$m and 2.323 $\mu$m in $K$, which trace the kinematics of young stars.  When available, we compare the stellar kinematics to those of various emission lines: \feii~1.644 $\mu$m, \brg~2.16 $\mu$m,  \molhy~2.12 $\mu$m.  For some galaxies, stellar kinematics were not available because the CO bandheads are redshifted out of $K$, and we rely on the aforementioned emission lines for this analysis.  Total exposure times, observed filters, and parameters of the observed galaxies are listed in Table~\ref{tbl:observingparams}.  Some of these data were presented in previous papers \citep{medling11, mrk273, nucleardisks}, these are indicated in Table~\ref{tbl:observingparams} as well.

We reduced our Keck II observations using the OSIRIS Data Reduction Pipeline\footnote{Available at \url{http://irlab.astro.ucla.edu/osiris/pipeline.html}.} version 2.3 (using the updated OSIRIS wavelength solution for data taken after October 2009) and our Keck I observations using version 3.  This pipeline includes modules to subtract sky frames, adjust channel levels, remove crosstalk, identify glitches, clean cosmic rays, extract a spectrum for each spatial pixel, assemble the spectra into a data cube, correct for atmospheric dispersion, perform telluric corrections, and mosaic frames together.  For some cases we utilized the Scaled Sky Subtraction module based on the technique outlined in \citet{Davies07}, which scales the thermal continuum and OH line groups separately to provide optimal sky subtraction; we modified the module to include a smoother subtraction of the thermal continuum, as described in \cite{nucleardisks}.

We imaged the tip-tilt stars for each galaxy during the course of the observations in order to obtain an estimate of the point-spread function (PSF).  A Moffat function was fit to each tip-tilt star and then broadened according to the distance between the target and the tip-tilt star.  This broadening accounts for the fact that the laser spot is not at infinity; therefore it only probes most of the turbulence that affects the target.  As a result, the further a target is from its tip-tilt star, the worse its PSF will be.  The isokinetic angle for the Keck II AO system, at which the Strehl ratio is reduced to 37\% of its peak value, is approximately 75 arcseconds.  We also use this an estimate of the Keck I AO system performance, which has not yet been characterized.

 \begin{deluxetable}{lcccccccc}
    \centering
    \tabletypesize{\scriptsize}
    \tablewidth{0pt}
    \tablecolumns{7}
    \tablecaption{Details of Observations}
    \tablehead{   
      \colhead{Galaxy Name} &
      \colhead{Redshift} &
      \colhead{Pixel Scale} &
      \colhead{UT Date(s)} &
      \colhead{Filter} &
      \colhead{Exp Time on} &
      \colhead{Exp Time on}  \\
      \colhead{} &
      \colhead{} &
      \colhead{(pc / 0.035\arcsec)} &
      \colhead{} &
      \colhead{} &
      \colhead{Target (minutes)} &
      \colhead{Sky (minutes)} & 
      	}
    \startdata
    CGCG436-030 & 0.0315 & 21.8 & 2012 Sep 30 & Kn5 & 30 & 5 \\
    IRASF01364-1042 & 0.0493 & 33.6 &  2012 Oct 01 & Kbb & 20 & 10 \\
    IIIZw035\tablenotemark{a} & 0.0278 & 19.5 & 2011 Dec 10 & Kbb  & 100 & 50 \\
    MCG+08-11-002 & 0.0195 & 13.9 & 2012 Jan 02 & Kbb  & 90 & 50 \\
    NGC~2623 & 0.0196 & 13.9 & 2010 Mar 04, 2010 Mar 05 & Kn5  & 100 & 50 \\
    UGC5101\tablenotemark{a} & 0.0413 & 27.0 & 2010 Mar 04, 2010 Mar 05 & Kn5  & 80 & 50 \\
    Mrk273/UGC8696\tablenotemark{b} & 0.038 & 26.4 & 2012 May 22 & Hn4  & 60 & 30  \\
    NGC~6240 N\tablenotemark{a} & 0.0244 & 17.2 &   2009 Jun 17 & Kn5  & 210 & 75  \\
    NGC~6240 S\tablenotemark{c} & 0.0244 & 17.2 & 2007 Apr 21 & Kn5  & 20 & 10  \\
    IRASF17207-0014\tablenotemark{a} & 0.0462 & 29.8 & 2011 May 23, 2011 May 24 &  Hn4  & 40 & 20  \\
    \enddata
    \tablenotetext{a}{Originally presented in \citet{nucleardisks}}
    \tablenotetext{b}{Originally presented in \citet{mrk273}}
    \tablenotetext{c}{Originally presented in \citet{medling11}}
    \label{tbl:observingparams}
  \end{deluxetable}


\section{Line Fitting}
\label{kinematicfitting}

We measured the kinematics following the methods presented in detail in \citet[for stellar kinematics]{medling11} and \citet[for emission line kinematics]{mrk273}, and reviewed again in \citet{nucleardisks}.  We briefly describe these techniques here.

For each measurement, we first calculate the signal-to-noise ratio in each pixel and bin them using optimal Voronoi tesselations \citep{voronoi} in order to require a certain signal-to-noise ratio for reliable measurements.  We used thresholds of 20 and 3 per resolution element for stellar kinematics and emission line kinematics, respectively.  Once binned, stellar kinematics were fit using the penalized pixel fitting routine\footnote{Available at \url{http://www-astro.physics.ox.ac.uk/~mxc/idl/}.} \citep{pPXF} and using $K$-band templates of late-type giants and supergiants from GNIRS \citep{GNIRS}.  Emission lines were fit with Gaussian profiles to determine the flux, velocity, and velocity dispersion in each bin.  For emission lines where multiple lines exist in a specific band (the 5 \molhy~transitions in the $K$-band, or \brd~and \brg), those lines were fit simultaneously, requiring that the velocity and velocity dispersion be consistent between lines.


\section{Black Hole Masses}
\label{results}

Our black hole mass measurement techniques have been demonstrated in two prior papers.  In \citet{medling11}, we place limits on the mass of the black hole in the south nucleus of NGC~6240 using 2-D stellar kinematic maps and two methods.  We calculate the lower limit to the mass by assuming the stars lie in a thin Keplerian disk, and fit a density profile including a black hole and a smoothly-varying spheroidal mass component: $\rho(r) = M_{BH} + \rho_{0} r^{-\gamma}$.  Because of the high spatial resolution of our data, we probe radii small enough that the first term dominates over the second.  The measured black hole mass here is a lower limit because this method ignores velocity dispersion (assuming that all measured velocity dispersion is due to material along the line of sight, not intrinsic to the disk).  Any intrinsic velocity dispersion in the disk will be indicative of extra mass in the black hole.  In the second scenario, we obtain an upper limit on the black hole mass by assuming the opposite: that all measured velocity dispersion is intrinsic to the disk.  Because the system is unrelaxed, there is unvirialized material along the line of sight increasing the velocity dispersion but not knowing about the central black hole mass.  To find this limit, we use Jeans Axisymmetric Mass models \citep[JAM;][]{JAM}.  We use both methods here for all galaxies with measured stellar kinematics, except for the north black hole in NGC~6240.  This nucleus contains patchy dust that prevents a consistent light profile from being produced by the JAM models; in order to use this measurement technique on that black hole, equivalently high resolution imaging at a longer wavelength ($L$-band or beyond) is required.

In \citet{mrk273}, we measure the mass of the black hole in the north nucleus of Mrk273 in a similar manner.  Since no stellar kinematics were available, the JAM models were not feasible.  When only gas kinematics are available, we model the gas as a thin Keplerian disk and proceed as in the first case above.  The black hole mass measured for this galaxy agrees well with the measurement made using the kinematics of an OH maser \citep{Klockner04}, and therefore provides independent justification for this technique.  We note that the thin Keplerian disk approximation provides reasonable $M_{BH}$ agreement to the maser measurement even though the disks analyzed in \citet{nucleardisks} are thick rather than thin, with $v/\sigma\sim$1-4.  We note that this thickness may cause our black hole mass measurements to underestimate the true black hole mass by up to $\sim25$\%.

With this approach, we are measuring the unresolved central mass within $\sim 25$ parsecs.  This is likely dominated by the black hole, and we therefore refer to it as such.  However, this mass includes the accretion disk of the black hole and may include a reservoir of gas feeding the accretion disk.  If the accretion disk is fueled smoothly by the nuclear disk, this mass will be accounted for by the radial profile; we cannot account for a pile-up of gas below our resolution limit.  The possible implications for this will be discussed further in Section~\ref{scalingrelations}.

In each case, errors on the black hole mass measurement were obtained using a Monte Carlo approach.  We created a model datacube and added random noise appropriate to the signal-to-noise of each spectrum and then refit the black hole mass for 100 iterations.  The width of the resulting mass distribution is the error on the measurement.

We have performed the above analysis for 9 additional black holes, listed in Table~\ref{tbl:observingparams}, and include the results in Table~\ref{tbl:mbh} and the fits in Appendix~\ref{BHmodelfigs}.  As in previous papers, these methods produce kinematic models in reasonable agreement with observations for most cases.  The ability to measure black hole masses using different tracers, and additionally to use bracketing assumptions about velocity dispersion for stellar models, improves the robustness of these results.

The numerical results from our black hole mass measurements are listed in Table~\ref{tbl:mbh}, along with parameters relevant to black hole scaling relations, pulled from the literature: the stellar velocity dispersion of the bulge, the stellar mass of the bulge, and the total stellar mass of the galaxy.  Though consistent treatment of such parameters when drawing from different literature sources is difficult, care was taken to select measurements most appropriate to the scaling relation.  The bulge luminosities and total stellar masses were taken from the GOALS papers \cite{Haan11} and \cite{U_SED}, respectively.  \citet{Haan11} measured the bulge luminosities using GALFIT \citep{Peng02, Peng10} to decompose $H$-band NICMOS images into multiple S\'ersic components.  These galaxies are morphologically complex, which can produce an increased uncertainty in bulge fit parameters; thus, the errors in bulge luminosities reported by \citet{Haan11} and reproduced in Table~\ref{tbl:mbh} may be larger than the equivalent measurements from Hubble imaging of isolated galaxies.  It is worth considering that we use the formal statistical errors produced by GALFIT; the analysis of \citet{Haussler07} has shown that these underestimate the true errors, which can contain contributions from profile mismatch and nearby neighbors, both relevant to our sample.  However, \citet{Haussler07} also find that GALFIT shows no systematic offset to fitting bulge luminosities of bright objects and that it is able to appropriately handle contamination by neighbors when fit simultaneously \citep[as was done in][]{Haan11}.  Still, for a more complete understanding of the errors associated with each bulge luminosity measurement, we refer the readers to the models and residuals for each system of \citet{Haan11}, published online in Figure Set 7.  Our total stellar masses were measured from SEDs using photometry masks \citep{U_SED} designed to incorporate all of the light from the galaxy.  We are thus not underestimating the total stellar mass, as can happen with fixed-aperture photometry, and we avoid potential biases related to varying host galaxy effective radius \citep{Hopkins07a,Hopkins07b,Hopkins09,Beifiori12}.
The GOALS survey papers allow us to be confident that the bulge luminosities and total stellar masses were measured consistently across our sample.  However, no such survey paper exists currently to measure the stellar velocity dispersions of the bulges.  Instead, we selected the stellar velocity dispersion measurement from the aperture that most closely aligned with the size of the bulge.  Unfortunately, no bulge information for MCG+08-11-002 is available in the literature.  Although it cannot be placed on scaling relations at this time, we include our black hole mass measurements for completeness.

 \begin{deluxetable}{lccccccc}
    \centering
    \tabletypesize{\scriptsize}
    \tablewidth{0pt}
    \tablecolumns{8}
    \tablecaption{Measured Black Hole Masses}
    \tablehead{   
      \colhead{Galaxy Name} &
      \colhead{Tracer\tablenotemark{a}} &
      \colhead{$M_{BH}$} &
      \colhead{$\sigma_{*,bulge}$ }  &
      \colhead{log($\frac{L_{H,bulge}}{L_{\sun}})$} &
      \colhead{$M_{*,total}$} &
      \colhead{Galaxy}  & \\
      \colhead{}&
      \colhead{}&
      \colhead{$(M_{\sun})$}&
      \colhead{(km s$^{-1}$)}&
      \colhead{}&
      \colhead{$(M_{\sun})$}&
      \colhead{References\tablenotemark{b}} &
      	}
    \startdata
    CGCG436-030 & \brg & $4.59 \substack{+0.52\\-0.48} \times 10^{8} $& $175\pm9$& $10.84\pm9.4$& $5.5\times 10^{9} $& 1,2,3  \\
    IRASF01364-1042 & \brg & $2.37 \substack{+0.01\\-0.1} \times 10^{9}$& $103\pm12$& $10.42\pm8.39$&$5.6 \times 10^{10}$ & 1,2,3 \\
     & \molhy & $2.12 \substack{+0.06\\-0.14} \times 10^{9}$ & &  \\
    IIIZw035 & stars - disk & $>6.8\substack{+0.1\\-4.0}\times 10^8 $& ... & $11.15\pm9.11$& $1.8\times10^{10}$&1,2  \\ 
[0.5ex]      & stars - JAM & $<2.0\substack{+0.5\\-0.7} \times 10^9 $& & & \\ 
     &\molhy& $2.59 \substack{+0.03 \\ -0.1} \times 10^{8} $& & & \\ 
[0.5ex]      & \brg & $3.4 \substack{+0.4 \\ -0.6} \times10^{8}$ & & & \\ 
[0.5ex] 
    MCG+08-11-002 & stars - disk & $>8.7 \substack{+3.1\\-3.0} \times 10^{7}$& ... & ... & ... & ... \\
     & stars - JAM  &$ <5.9\substack{+7.1\\-1.6}\times 10^{8} $ & & & \\
     & \molhy  & $6.9 \substack{+1.0\\-1.6} \times 10^{7}$ & & & \\
     & \brg  & $2.3\substack{+1.2\\-0.9} \times 10^7$ & & & \\
[0.5ex]     NGC~2623 & stars - disk & $>2.9\substack{+0.1\\-0.7}\times10^8$& $95\pm13$& $10.61\pm8.57$ & $2.4\times10^{10}$& 1,2,4 \\
[0.5ex]      & stars - JAM & $<4.7\substack{+1.2 \\ -2.6}\times10^8 $& & &\\
[0.5ex]     UGC5101 & stars - disk & $>6.5 \substack{ +3.5\\-2.1} \times 10^8$& $287\pm11$ & $11.58\pm9.75$& $1.7\times10^{11}$& 1,2,5  \\
     & stars - JAM & $<5.4\substack{+70.8\\-0.4} \times10^8$& & &\\
[0.5ex]     Mrk273 N &\feii & $1.0\pm 0.1\times10^9$ & $285\pm30$& ...& $6.9\times10^{11}$& 1,6  \\
[0.5ex]     NGC~6240 N\tablenotemark{c} & stars - disk &$ > 8.8\substack{+0.7 \\-0.1 } \times10^8$  & $174\pm54$& $10.81\pm9.38$& $3.9\times10^{11}$&2,7,8  \\
[0.5ex]     NGC~6240 S & stars - disk & $>8.7\pm0.3\times10^8$ & $236\pm24$& $11.29\pm10.25$& $3.9\times10^{11}$&2,7,8 \\
[0.5ex]      & stars - JAM & $<2.0\pm0.2\times10^9$  & & &  \\
    IRASF17207-0014 E & \feii & $2.5\substack{+0.03\\-0.3}\times10^9$ & $229\pm15$&$10.39\pm9.06$ & $1.52\times10^{11}$ & 2,6,7  \\
    IRASF17207-0014 W & \feii & $8.9\substack{+4.0\\-1.6}\times10^{7}$& $229\pm15$ & $10.39\pm9.06$& $1.52\times10^{11}$ & 2,6,7 \\
    \enddata
    \tablenotetext{a}{All measurements with gas tracers use the thin disk method, those with stellar tracers use the disk or JAM method as marked.}
    \tablenotetext{b}{Galaxy parameters in columns 4-6 taken from the following references: 1 - \cite{U_SED}, 2 - \cite{Haan11},  3 - \cite{Tremonti04}, 4 - \cite{Shier96}, 5 - \cite{Rothberg06}, 6 - \cite{Dasyra06}, 7 - \cite{Howell10}, 8 - \cite{Tecza}} 
    \tablenotetext{c}{Due to extensive and patchy dust coverage, a JAM model could not be completed for NGC~6240N without additional longer-wavelength ($L$-band or beyond) imaging data.}
    \label{tbl:mbh}
  \end{deluxetable}
  \clearpage

\section{Mergers and Black Hole Growth}
\label{scalingrelations}
\subsection{Placing Black Holes on Scaling Relations}

Using our measured black hole masses and host galaxy parameters from the literature, all listed in Table~\ref{tbl:mbh}, we place the galaxies on several black hole scaling relations.  

\subsubsection{The $M_{BH} - \sigma_{*}$ Relation}

\begin{figure}[ht]
\includegraphics[trim=1cm 0cm 0cm 0cm, scale=.7]{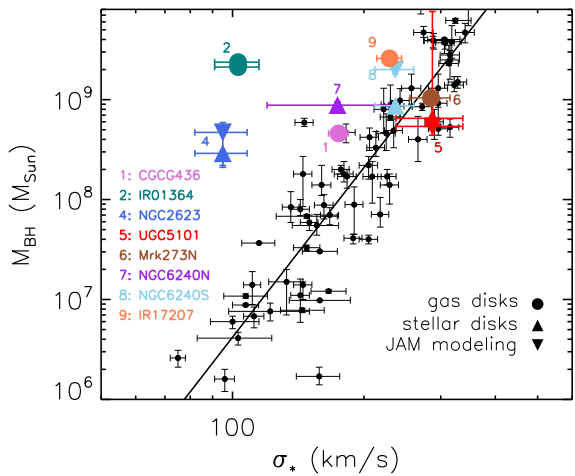}
\caption[$M_{BH} - \sigma_{*}$ Relation ]{The $M_{BH} - \sigma_{*}$ relation for isolated galaxies from \cite{McConnell13} (black data and best-fit line: log($M_{BH}/M_{\sun}) = 8.32 + 5.64$ log$(\sigma_{*}/200$ km/s) with our merging galaxies marked in large colored points.  Symbols indicate which method was used for that mass determination: gas disk modeling in circles, stellar disk modeling in upward triangles, and JAM modeling in downward triangles.  Stellar disk models represent lower limits to the black hole mass, while JAM models represent upper limits; see text for details.  Mrk273 N's point (6; brown) is the mass measurement for only the north black hole, though there is a second in the system \citep[see][]{mrk273}.
}
\label{Msigma}
\end{figure}

In Figure~\ref{Msigma}, we show the $M_{BH} - \sigma_{*}$ relation from \cite{McConnell13} with our points overlaid.  

The black holes in our sample lie either within the scatter of the relation or above it.  Although a particular galaxy may fall within the scatter of the relation, we want instead to consider whether our sample as a whole is behaving as predicted by the $M_{BH} - \sigma_{*}$ relation.  We calculate the distances from the $M_{BH} - \sigma_{*}$ relation for both our dataset and the reference population, and then perform a two-sided Kolmogorov-Smirnoff test on the two populations.  We find a p-value of 0.003, indicating that the chance of these two being drawn from the same population is less than 1\%.  To mitigate the possible effects of outliers in our small sample, we use the balanced bootstrap method\footnote{Balanced bootstrap resampling is similar to ordinary bootstrap resampling but requires that each observed value appears with equal frequency in the resampled data \citep[e.g.][]{Gleason88}.  We used the IDL routine BBOOTSTRAP, available at \url{http://www.astro.washington.edu/docs/idl/cgi-bin/getpro/library14.html?BBOOTSTRAP}.} to estimate the average offset from the relation.  This reveals an average offset of $0.15\pm0.06$ in plot units; we thus consider this offset a 2.6-sigma result.

Our sample of black holes are thus, on average, above the relation, suggesting with moderate significance that black holes may grow in mergers before the bulges are virialized.  Measuring the velocity dispersion of stars in the bulge can be difficult to define when two bulges are in the process of merging.  This variation in measured velocity dispersion (both due to the process of merging and due to geometric variations based on the line-of-sight) has been quantified in the hydrodynamical galaxy merger simulations of \citet{Stickley14}.  Their analysis shows a maximum predicted mismeasurement in velocity dispersion of approximately 50\%, which is insufficient to bring all our systems on to the relation.

\subsubsection{The $M_{BH} - L_{H,bulge}$ Relation}

\begin{figure}[ht]
\includegraphics[trim=1cm 0cm 0cm 0cm, scale=.7]{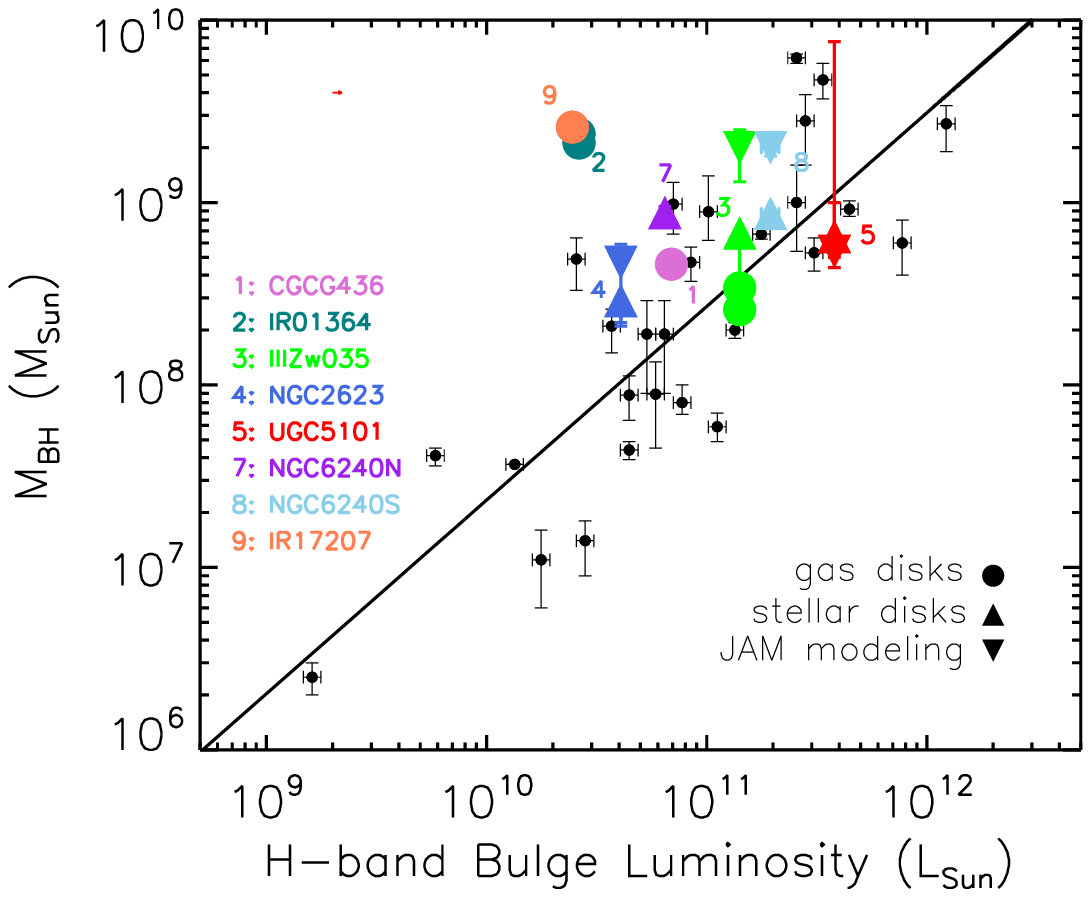}
\caption[$M_{BH} - L_{H,bulge}$ Relation]{The $M_{BH} - L_{H,bulge}$ relation.  The black points have luminosities taken from \cite{MarconiHunt03} and updated black hole masses from \cite{McConnell13}.  The solid line is a log-log fit to the updated data: $log(\frac{M_{BH}}{M_{\sun}}) = 8.22 + 1.06~log(\frac{L_{H,bulge}}{10^{10.8} L_{\sun}})$.  Galaxies presented here are indicated by large colored points, with parameters listed in Table~\ref{tbl:mbh}.  The bulge luminosities were not corrected for extinction; the arrow in the top left shows the typical magnitude of such a correction (10\%).  Symbols indicate which method was used for that mass determination: gas disk modeling in circles, stellar disk modeling in upward triangles, and JAM modeling in downward triangles.  Stellar disk models represent lower limits to the black hole mass, while JAM models represent upper limits; see text for details.  
}
\label{Mbulgelum}
\end{figure}

In Figure~\ref{Mbulgelum}, we show the $M_{BH} - L_{H,bulge}$ relation.  The background galaxies were plotted using $L_{H,bulge}$ from \cite{MarconiHunt03}, with updated black hole masses from \cite{McConnell13}.  We have updated the best-fit line with these numbers, producing the relation:
\begin{displaymath}
log(\frac{M_{BH}}{M_{\sun}}) = 8.22 + 1.06~log(\frac{L_{H,bulge}}{10^{10.8} L_{\sun}})
\end{displaymath}

The black holes from our sample fall either within the scatter or above the $M_{BH} - L_{H,bulge}$ relation.  As with the $M_{BH} - \sigma_{*}$  relation, we perform a two-sided Kolmogorov-Smirnoff test to determine if our sample as a whole falls significantly above the relation; we find a p-value of $0.018$, indicating that these black hole masses are unlikely to be pulled from the sample distribution as the reference sample.  We also calculate the mean offset using the balanced bootstrap method, $0.47\pm0.14$ in plot units, showing an offset with $\sim3.4$-sigma significance.

We note two possible calibration effects when considering our bulge luminosity measurements.  First, although \citet{MarconiHunt03} did not correct their bulge luminosities for internal extinction, we consider the effects such a correction would have on our sample because our galaxies are quite dusty.  Several of our galaxies have $JHK$ imaging published by \citet{Scoville00}, enabling a direct estimate of typical extinction over the central few kiloparsecs.  These calculations estimate that the intrinsic luminosity of the bulges in our sample may be $\sim10$\% higher.  This correction (indicated by red arrow in the top left of Figure~\ref{Mbulgelum}) is smaller than the size of our points and can be safely ignored.  Second, we investigate whether the varying bulge S\'ersic indices of the fits in \citet{Haan11} may be biasing our luminosity estimates.  When fixing $n$ to 4, the bulge luminosities of several of our galaxies change significantly, some higher and some lower (S. Haan, private communication).  The net effect with this change is less scatter between our galaxies: a smaller but more statistically significant offset.  We conclude that, while the S\'ersic index fit may have a strong effect on an individual system's predicted black hole mass when using the $M_{BH} - L_{H,bulge}$ relation, the offset in our data is not caused by the fitting technique of these bulge luminosities.  We also note that the two galaxies which fall furthest above the relation, IRASF01364-1042 and IRASF17207-0014, do not show atypically large residuals in the GALFIT bulge fits (indeed, the residuals of the former appear smaller than average for this sample), which suggests that their offset is not due to errors in the bulge luminosities of \citet{Haan11}.

Again, our sample of merging galaxies falls above this relation, suggesting that either stars have not been forming in the bulge as quickly as the black holes have been growing, or that a large mass of stars is still strewn about in tidal tails and unrelaxed features that will eventually become part of the bulge.  
As the $H$-band traces older populations of stars, it may be that some stars have recently formed in the bulge, but that this measurement misses them.  Although $M_{BH} - L_{bulge}$ relations do exist for bluer colors, we are unfortunately unable to utilize them for this sample because the large amounts of dust obscure the galaxy cores, making bulge luminosity measurements in the optical highly uncertain.  We have compared our bulge luminosities to those fit by \citet{Kim13} in the $I$-band, and found that (for typical mass-to-light ratios) our $H$-band bulges are more massive.  This simply confirms that dust is a bigger problem even in the $I$-band than missing stellar populations are in the $H$-band for this sample.

\subsubsection{The $M_{BH} - M_{*}$ Relation}

\begin{figure} 
\includegraphics[trim=1cm 0cm 0cm 0cm, scale=.7]{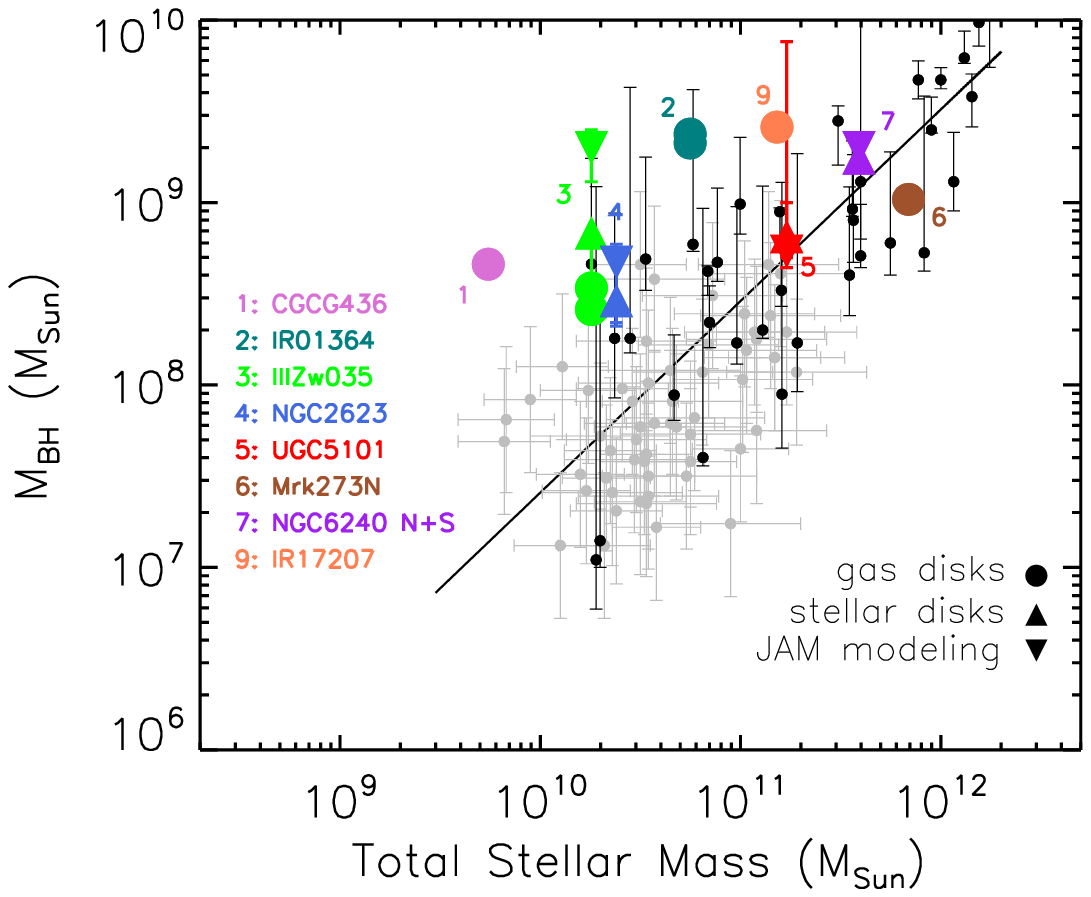}
\caption[$M_{BH} - M_{*}$ relation]{The $M_{BH} - M_{*}$ relation.  The grey points have luminosities and total stellar masses from \cite{Bennert11} and \cite{Cisternas11}, who suggest that higher redshift galaxies sit on the local $M_{BH} - M_{*,bulge}$ relation if you include total stellar mass instead of only bulge mass.  In this way, they attribute evolution in the black hole scaling relation to the evolution of the bulge/disk fraction with redshift.  We include also black points for local elliptical galaxies and the fit to for $M_{BH} - M_{*,bulge}$ from \cite[][log($M_{BH}/M_{\sun}) = 8.46 + 1.05$ log($M_{*}/ 10^{11} M_{\sun}$]{McConnell13}.  Merging galaxies from this work are indicated by large colored points, with parameters listed in Table~\ref{tbl:mbh}.  Symbols indicate which method was used for that mass determination: gas disk modeling in circles, stellar disk modeling in upward triangles, and JAM modeling in downward triangles.  Stellar disk models represent lower limits to the black hole mass, while JAM models represent upper limits; see text for details.  NGC~6240's two nuclei were summed together and appear as one point (7; dark purple) on this plot, paired with the entire system's total stellar mass.  Mrk273 N's point (6; brown) is the mass measurement for only one black hole, though there is a second in the system \citep[see][]{mrk273}.
}
\label{Mtotalmass}
\end{figure}

In Figure~\ref{Mtotalmass}, we plot $M_{BH} - M_{*}$ using points from \cite{Bennert11} and \cite{Cisternas11}.  These authors suggest that higher redshift galaxies sit on the local $M_{BH} - M_{*,bulge}$ relation if you include total stellar mass instead of only bulge mass.  In this way, they attribute evolution in the black hole scaling relation to the evolution of the fraction of stellar mass in the bulge.  We include as a solid line the updated fit to the $M_{BH} - M_{*,bulge}$ from \cite{McConnell13}, and plot our black hole masses in the same colored points as in Figures~\ref{Msigma} and \ref{Mbulgelum}.  

To determine if this sample of systems is in agreement with the $M_{BH} - M_{*}$ relation, we perform a similar two-sided Kolmogorov-Smirnoff statistical test as above.  We find that, compared to the local elliptical galaxies of \cite{McConnell13}, the p-value obtained is only 0.01, less significant than above.  When comparing to the higher-redshift samples of total stellar mass, we obtain a p-value of 0.0008, and to the combined sample, a p-value of 0.004.  The balanced bootstrap method finds an offset from the McConnell \& Ma relation of $0.46\pm0.14$ in plot units, similar to the $M_{BH} - L_{bulge}$ relation.

We thus conclude that our sample of galaxies on average lies above this scaling relation as well.  We also note that the one point that falls below the relation, the brown circle Mrk273 N (labeled 6), represents the mass in only one out of two black holes and is therefore artificially low.  Because this relation takes into account the entirety of stellar mass, the fact that many of our points fall above the relation suggests that star formation as a whole is not outpacing black hole growth during these mergers.

\subsubsection{Overall Significance of Scaling Relation Offset}

In the previous sections, we have compared our galaxies with the $M_{BH} - \sigma_{*}$ and $M_{BH} - M_{*,bulge}$ relations from \citet{McConnell13}.  We have also compared them to those from \citet{KormendyHo13}, and find that our offset remains.

Two galaxies appear to drive the offset in the $M_{BH} - \sigma_{*}$ relation: IRASF01364-1042 and NGC~2623.  Although we note that these are not outliers in a statistical sense, we must be certain that they are not unfairly biasing our results.  Because our sample for each scaling relation has only 8 points, throwing out these two galaxies has a significant effect in all cases through reducing the sample size.  We note, however: when removing the two galaxies from the $M_{BH} - \sigma_{*}$ relation, the measured offset decreases quite a bit, while the uncertainty in the offset also increases (producing only a 1.9-$\sigma$ result).  For the other two relations, the measured offset decreases only slightly, and the main effect is an increase in uncertainty (so the significance still decreases from 3.3-$\sigma$ to 2.3-$\sigma$).  This demonstrates that the offsets aren't driven by these two galaxies in the $M_{BH} - L_{bulge}$ and $M_{BH} - M_{*}$ relations.

Due to the small sizes of our samples, the appropriate way to deal with potential outliers is to use the balanced bootstrap test, which we have done.  The balanced bootstrap performs hundreds of resamplings in order to estimate the significance of a result, giving equal weight to all measurements.  (Non-`balanced' bootstrapping may resample some points more than others, which could cause a bias when one or more potential outliers exist.  By balancing the bootstrapping, we force the equal treatment of all measurements; this prevents a single point (or two) from biasing the results.)  By doing so, we are able to handle outliers (or near-outliers) without diminishing our sample size.

The black holes in this sample of gas-rich major mergers consistently fall on or above the scaling relations for passive galaxies with stellar velocity dispersion, bulge luminosity and total stellar mass.  Although a number of our galaxies lie within the scatter of these scaling relations, our statistical tests show they are significantly more likely to fall on the upper side of the scatter than randomly distributed about the relations.  Though the offset from the $M_{BH} - \sigma_{*}$ relation is only a 2.6-sigma result, the agreement with the two other significant relations lends it confidence.  Taken together, these results show that increased black hole growth in these mergers has already begun and may be proceeding more rapidly than growth of the host galaxy.  

Although we have a relatively small sample of black holes, it is striking that we already see a significant offset.  As more black hole masses are measured in gas-rich mergers, the larger sample will provide a more accurate picture of how offset such systems are from black hole scaling relations.

We caution the reader against considering the distance from a scaling relation as a direct proxy for merger stage.  This is clearly evident when comparing where a system falls on one scaling relation versus another.  For example, IRASF17207-0014 falls closer to the $M_{BH} - \sigma_{*}$ relation than to the $M_{BH} - L_{H,bulge}$ relation; for NGC~6240S, the reverse is true.  This discrepancy could be caused by at least two possible mechanisms: 1) The host galaxy parameters trace different aspects of galaxy growth, which may evolve differently.  For example, the bulge velocity dispersion traces a convolution of mass growth with the virialization timescale whereas the bulge luminosity evolution may rely less stringently on the dynamical relaxation time.  2) As discussed above, the simulations of \citet{Stickley14} show that the measured stellar velocity dispersion of similar mergers oscillates during the course of the merger.  Analogous simulations would have to be performed for bulge luminosity and total stellar mass measurements to determine if they ought to oscillate as well.  Even if all three quantities were expected to oscillate during a merger, it seems contrived to require that they oscillate in lockstep.  To determine precisely how galaxies evolve along a scaling relation (and if they evolve the same way along each scaling relation), a larger sample of systems is required.  By binning over merger stage, the range of distances from scaling relations can be seen, and key parameters affecting this can be identified.  We leave this analysis for future studies.

\subsection{Interpreting Overmassive Black Holes}

In this section, we discuss several plausibility arguments related to the above results.  We consider the growth timescales and the available fuel for both black holes and the host galaxies.

\subsubsection{Has There Been Enough Time for these Black Holes to Grow?}

We posit that these systems followed black hole scaling relations before the merger began and that, since then, mass has been funneled to the center to enable black hole growth.  For simplicity, we assume that each galaxy previously fell precisely on every scaling relation; of course, as each relation contains intrinsic scatter, this is unlikely.  However, we have no reason to believe that these systems were systematically offset from scaling relations before the mergers began, and thus this simplification should average out in the order-of-magnitude calculations of this and the following sections.

With that assumption, we consider how long it would take for a black hole, growing at the Eddington rate, to reach this far beyond its original mass.  In Table~\ref{tbl:eddrate}, we calculate this for our sample of galaxies.  We see that the required times for these black holes to grow, while accreting at the Eddington rate, are a few tens to hundreds of millions of years.  As this is less than or comparable to a merger timescale, it is reasonable that these systems could have lain on scaling relations before the merger, and risen only after the merger began.  As black hole growth is known to be episodic, we note that these systems could often have been accreting at sub-Eddington rates for timescales less than those noted, and that the accretion times need not have been continuous.

The merger of one of these objects, NGC~2623, has been modeled by \citet{Privon13} using H I kinematics from the Very Large Array.  Their dynamical models indicate that this object is $2.2 \times 10^{8}$ years past first pericenter passage, within the range of growth timescales we see for this object in Table~\ref{tbl:eddrate} ($0.6-2.3 \times10^{8}$ years).

We also repeat that our black hole mass measurements produce dynamical masses of a central point source, which to our spatial resolution could be anything smaller than $\sim25$pc.  It is therefore plausible that some of the mass measured is in the accretion disk (or, perhaps, in a nearby reservoir feeding the accretion disk).  If that's the case, any mass that has not accreted directly onto the black hole would not cause the black hole to turn on as an AGN.  In such a scenario, the Eddington rate may not be a relevant limiting factor in this growth phase.

 \begin{deluxetable}{lcccccccc}
    \centering
    \tabletypesize{\scriptsize}
    \tablewidth{0pt}
    \tablecolumns{8}
    \tablecaption{Black Hole Growth Timescales}
    \tablehead{   
     \colhead{} &
     \colhead{} &
\multicolumn{2}{c}{From $\sigma_{*,bulge}$ } &
\multicolumn{2}{c}{From $L_{H,bulge}$ } &
\multicolumn{2}{c}{From $M_{*,total}$ } &
      \\
      \cline{3-4}  \cline{7-8} \\
      \colhead{Galaxy Name} & 
      \colhead{Mean} &
      \colhead{Implied Mass} & 
      \colhead{Time at} &
      \colhead{Implied Mass} & 
      \colhead{Time at} &
      \colhead{Implied Mass} & 
      \colhead{Time at} &
      \\
      \colhead{} & 
      \colhead{$M_{BH}$ (M$_{\sun}$)} &
      \colhead{Growth (M$_{\sun}$)} & 
      \colhead{Edd (yr)} &
      \colhead{Growth (M$_{\sun}$)} & 
      \colhead{Edd(yr)} &
      \colhead{Growth (M$_{\sun}$)} & 
      \colhead{Edd (yr)} &
      	}
    \startdata
CGCG436-030 & 4.6e8 & 3.6e8 & 7.7e7 &  2.8e8 & 4.6e7 & 4.5e8 & 1.8e8  \\
IRASF01364-1042 & 2.3e9 & 2.3e9 & 3.1e8 &  2.2e9 & 1.8e8 & 2.2e9 & 1.3e8   \\
IIIZw035 & 2.6e8 & & & -- & -- & 2.1e8 & 8.5e7   \\
NGC~2623 & 3.0e8 & 2.9e8 & 2.3e8 & 1.9e8 & 5.3e7 &  2.3e8 & 7.6e7  \\
UGC5101 & 5.5e8 & -- & -- & -- & -- & 4.9e7 & 4.6e6  \\
Mrk273 N & 1.0e9 & -- & -- & & & -- & --  \\
\vspace{-1.5mm} NGC~6240N & 8.8e8 & 7.8e8 & 1.1e8 & 7.1e8 & 8.2e7 &     \\
\vspace{-1.5mm} & & & & & & 5.6e8 & 1.9e7 \\
NGC~6240S & 8.9e8 & 3.6e8 & 2.6e7 & 3.5e8 & 2.5e7 &     \\
IRASF17207-0014 & 2.6e9 & 2.1e9 & 8.8e7 & 2.5e9 & 1.9e8 &  2.1e9 & 8.8e7   \\
    \enddata
\tablecomments{Implied mass growth columns give the difference between the measured $M_{BH}$ and that predicted from the indicated host galaxy properties and scaling relations.  The time at Eddington rate columns are caculated from $M_{BH,now} / M_{BH,predicted} = e^{t/\tau_{Sal}}$, where $\tau_{Sal}$ is the Salpeter timescale of $5.7 \times 10^{7}$ years.  Columns with blank entries do not have measured host galaxy properties; entries marked with a -- indicate systems that lie on or below scaling relations, and therefore require zero growth.
}
    \label{tbl:eddrate}
  \end{deluxetable}

\subsubsection{How Long Would It Take To Return to Scaling Relations?}
\label{SFRsection}

The gas-rich major mergers we've studied here appear to have black holes significantly larger than would be expected based on their host galaxy properties.  We assume that these systems, post-merger, ought to fall on scaling relations.  As our systems are currently undergoing enhanced star formation, it is likely that the host galaxies will indeed continue to grow over the course of the merger, moving back towards the local scaling relations in Figures~\ref{Msigma}-\ref{Mtotalmass}.  In Table~\ref{tbl:SFR}, we show the relevant star formation rates from \citet{Howell10} and calculate how long it would take at this star formation rate to bring the total stellar masses up to those predicted by the measured black hole masses.

We see the typical timescale for these objects to reach predicted total stellar masses is 1-2 Gyr at the present star formation rate.  The dynamical models of \citet{Privon13} find merger timescales for their sample of 4 systems to range from 140 Myr - 1.4 Gyr, depending on the initial conditions of the merger.  Apart from IRASF10364-1042, our galaxies are not far from this range.  Though star formation is a variable quantity, it is plausible for these galaxies to return to the $M_{BH} - M_{*}$ relation by the time they have finished merging.

As noted above, it is possible that not all the mass we measured has been accreted by the black hole yet.  If this is the case, it allows for the possibility that some of this mass might be ejected via AGN-driven winds and may indeed never be accreted; we discuss this scenario in Section~\ref{outflows}.  If this were to happen, the stellar mass growth need not be so high to return these systems to the $M_{BH} - M_{*,total}$ relation.

 \begin{deluxetable}{lcccc}
    \centering
    \tabletypesize{\scriptsize}
    \tablewidth{0pt}
    \tablecolumns{4}
    \tablecaption{Star Formation Requirements}
    \tablehead{   
      \colhead{Galaxy Name} & 
      \colhead{$\Delta M_{*,total}$\tablenotemark{a}} &
      \colhead{SFR ($M_{\sun}$/yr)\tablenotemark{b}} & 
      \colhead{Time Required (yr)\tablenotemark{c}} &
      \colhead{Available $M_{H_{2}}$ ($M_{\sun}$)\tablenotemark{d}} 
      	}
    \startdata
CGCG436-030 & 1.5e11 & 85.87 & 1.7e9 & ...  \\
IRASF01364-1042 & 6.7e11 & 122.61 & 5.5e9 & ... \\
IIIZw035 & 7.3e10 & ... & ... & ...\\
NGC~2623 & 7.9e10 & 69.19 & 1.1e9 & 1.66e9 \\
UGC5101 &  1.6e10 & 180.18 & 8.7e7 & 2.65e9\\
Mrk273 N\tablenotemark{e} & -- & -- & -- & -- \\
NGC~6240 N+S & 1.7e11 & 148.44 & 1.2e9 & 1.18e10 \\
IRASF17207-0014 & 6.6e11 & 501.22 & 1.3e9  & 7.13e9 \\
    \enddata
\tablenotetext{a}{$\Delta M_{*,total}$ gives the difference between the current $M_{*,total}$ (column 6 in Table~\ref{tbl:mbh}) and that predicted from the current mean measured $M_{BH}$ and the relevant scaling relation.}
\tablenotetext{b}{Star formation rates taken from \citet{Howell10}.}
\tablenotetext{c}{Time required = $\Delta M_{*,total}$ / SFR}
\tablenotetext{d}{From \citet{Wilson08} when available}
\tablenotetext{e}{Mrk273 N already falls on the scaling relation, and therefore no additional star formation is needed.}
    \label{tbl:SFR}
  \end{deluxetable}

\subsubsection{Is There Enough Gas Mass Left to Grow the Host Galaxy?}

Regardless of the rate of star formation discussed in Section~\ref{SFRsection}, we might ask if there is enough molecular gas present in these galaxies to increase the $M_{*,total}$ sufficiently if all the gas were to be turned into stars eventually.  We include the available mass in cold gas measured for each of these galaxies, when available, using H${_2}$ masses measured by \citet{Wilson08}.  However, \citet{Wilson08} point out that the CO-to-\molhy~conversion factor used is a primary source of uncertainty, and caution that their masses could be underestimated by up to a factor of 5, and that AGN contamination in UGC5101 could cause further underestimation.  Taking these uncertainties into account, the gas masses could be sufficient in UGC5101 and NGC~6240, but perhaps not in NGC~2623 or IRASF17207-0014.

We note that this estimate ignores atomic gas which may eventually cool and form molecular gas and then stars; this quantity could be substantial, and that these systems could also continue to accrete gas from the circumgalactic medium.  However, a number of ULIRGs are also hosts to massive molecular outflows \citep{Rupke11,Sturm11,Spoon13,Veilleux13}, which could instead further deplete the gas.

Thus, the low gas mass estimates may support the possibility that not all excess mass measured from the central regions will eventually be accreted to the central black hole.

\subsection{Are These Masses Truly Measuring Black Holes?}
\label{outflows}

Our technique produces a dynamical measurement of the mass produced by a point source in the centers of these galaxies.  Though the most straightforward interpretation of this is that the mass measured describes a black hole mass, it is possible that the mass distribution merely appears as a point source to our spatial resolutions.  That is, anything significantly smaller than our pixel scales of $\sim25$pc, and above the disk's radial density profile, would appear as a black hole to our measurements.

As discussed in \citet{medling11} and \citet{mrk273}, it is possible that some mass may be due to a nuclear star cluster.  However, we rule out this interpretation for two reasons: 1) nuclear star clusters commonly have radii $\sim100$pc \citep{Murray09}, which would not appear as a point source to our dynamical modeling, and 2) the continuum luminosity present is not sufficient to account for a significant fraction of the central masses being due to stars.  (Full spectral energy distribution fitting to the star clusters in and around the nuclear regions of these galaxies will be presented in a future paper.)  We cannot, however, rule out the idea that gas present in the center could someday form a nuclear star cluster.

Galaxy merger simulations commonly assume Bondi-Hoyle accretion \citep{Bondi44, Bondi52, Springel05_feedback}, a prescription in which effectively gas particles which enter within the sphere of influence of the black hole are automatically accreted so long as the Eddington rate is not exceeded.  However, gas inflow likely has angular momentum which must be discarded before accretion can occur, which can delay the accretion event by e.g. the viscous timescale \citep{Power11,Wurster13}.  Though we don't have strong constraints on the properties of AGN accretion disks, estimates predict that viscosity could slow down accretion by more than a Hubble time \citep{King08}.  Although the existence of AGN confirms that accretion can occur on shorter timescales, the delay may not be negligible.  This allows for the possibility that gas can build up in the core of a galaxy around the black hole without having yet been accreted.  This accretion disk (and, possibly, a reservoir feeding the accretion disk) of gas could contain a substantial amount of mass en route to the black hole.  This mass would all appear as a point source to our dynamical models.  

Given the scaling arguments in the sections above, we also note the possibility that this putative accreting gas may never make it to the black hole.  Although the black hole dominates the gravitational pull on this gas, we find it plausible that when the black hole begins to accrete, its feedback may blow some of the rest of the gas away.  Thus, the extra mass that we see here could be the reservoir for massive outflows seen in a variety of local AGN \citep[e.g.][]{Alatalo11,Rupke11,Sturm11,Veilleux13,Spoon13}.  At mass outflow rates of tens or hundreds of solar masses per year, the excess mass could easily be ejected in the time remaining in the merger.  It is not currently known where the mass carried in these outflows is drawn from, or in particular, how much of it might come from within $\sim25$ pc.

Simulations have also suggested that nuclear gas disks may leave behind remnant stellar disks on 1-10 parsec scales with masses 0.1-1 $M_{BH}$ \citep{Hopkins10_andromedadisk}; if this were to occur, the final black hole masses in our sample will be decreased by 10-50\%.  We note, however, that the black hole masses from which the scaling relations were calibrated did not have sufficient resolution to separate out remnant disks on 1-10 parsec scales either; thus we only introduce a bias if our sample is more likely to host such disks than the typical galaxy.  We note that such remnant disks were discovered in galaxy merger simulations but that they did not form until after the final coalescence of the two nuclei.  Thus it is not clear whether we would expect such disks (broken from the larger nuclear disk on hundred parsec scales) to be present in our sample.

Though the above possibilities may moderate the black hole masses measured by attributing some mass to other features, we still conclude that the black holes (combined with their accretion disks) in this sample of gas-rich mergers do not lie below black hole scaling relations, and in many cases are significantly more massive than their host galaxy properties would predict.  This is directly opposite the predictions of quasar-mode feedback theories, in which black hole growth is delayed until the final stages of the galaxy merger when it cuts off star formation.  As we note that (U)LIRGs are prototypically expected to evolve into quasars \citep{Sanders96}, we find this as further evidence of some delay in feedback timescale during a merger.  Such a delay (due e.g. to a viscous accretion disk, as discussed above) could reconcile the late-stage AGN feedback paradigm with the early mass growth presented here.  Testing this scenario requires spatial resolutions higher by a factor of 10-20, which may be feasible in the upcoming era of thirty-meter-class telescopes.

\subsection{Comparison with Previous Findings}

These results differ from the conclusions drawn by \cite{KormendyHo13}, who show five merging galaxies falling below the $M_{BH} - \sigma_{*}$ and $M_{BH} - L_{K,bulge}$ relations.  They suggest that black hole growth might lag bulge formation in mergers, the opposite of the findings of this work.  It is important to note the differing samples when considering these discrepant results.  Our sample of merging galaxies contains major gas-rich mergers; the merging galaxies considered in \cite{KormendyHo13} are mainly minor mergers and/or gas-poor mergers.  It is not unreasonable to imagine that such different types of mergers would have different effects on their central black holes.  For example, a minor or gas-poor merger might use up all incoming new gas in star formation before it reaches the central black hole, causing a delay in black hole accretion until stellar feedback produces new gas to fuel it.  A major gas-rich merger may have sufficient gas inflow to enable both star formation and black hole growth to begin relatively early.

A sample of black holes in submillimeter galaxies appear to fall below the locally-defined $M_{BH} - M_{*,bulge}$ relation by 1-2 orders of magnitude \citep{Borys05,Alexander08}.  These galaxies are thought to be the high-redshift (z$\sim2$) counterparts of local ULIRGs and therefore might be reasonable to expect that they would behave in the same way as the sample presented here.  However, the discrepancy may in fact be due to their usage of broad line velocity dispersions to infer $M_{BH}$; indeed, when applying the same method in one local ULIRG, \citet{Alexander08} find rough agreement with their submillimeter galaxies.  It is also possible that differences between the samples such as Eddington ratio, redshift, and merger stage might affect the black hole growth timescales.


\section{Conclusions}
\label{conclusions}

We have presented high spatial resolution near-infrared integral field spectroscopy of nuclear disks in nearby merging (U)LIRGs.  These gas-rich mergers funnel gas to the galaxy centers where the gas forms disks; stars then form \textit{in situ}, creating a stellar disk as well.  These disks, which were described in detail in \citet{nucleardisks}, provide an alternative method to measuring black hole masses.  Three-integral orbital superposition models rely on an assumption of virialization and emission line black hole mass diagnostics require broad lines; neither of these methods is feasible in such systems.  However, with adaptive optics, we have produced two-dimensional kinematic maps at spatial resolutions close to or better than radius of the sphere of influence of the black hole.  Because the dynamical timescale this close to a black hole is short, the material quickly becomes virialized, and kinematic modeling is possible.

We implemented two different techniques using these kinematic maps, originally presented in \citet{medling11} and \citet{mrk273}, to measure 9 new black hole masses, producing a sample of black hole masses in 9 merging systems.  By obtaining both gas and stellar kinematics for many of these systems, we are able to provide multiple independent measurements for a large fraction of our sample.  Gaseous disks are modeled as thin Keplerian disks.  Stellar disks are more likely to have pressure support, but measuring this is not straightforward, so we place bracketing measurements using two assumptions.  To provide a lower limit to the black hole mass, we assume the stars are also in a thin Keplerian disk; any pressure support would increase the black hole mass.  We also use Jeans Axisymmetric Mass models to take into account the velocity dispersion; since this velocity dispersion may be inflated by intervening unvirialized material along the line of sight, this mass measurement is an upper limit.  

As the black hole masses measured here are in fact masses enclosed in $\sim25$pc, some of the mass attributed to black holes here may be gas that has not yet accreted onto the black hole.  Although this mass is gravitationally bound to the black holes, we note that it is possible that future AGN-driven outflows may carry a fraction of it away, thus diminishing the final black hole mass.  Though such outflows have now been seen in a number of (U)LIRGs, further studies will reveal how frequently they occur, how much mass they carry, and how much of the loaded mass comes from within $\sim25$pc.

We have placed these enclosed masses on three local black hole scaling relations, $M_{BH} - \sigma_{*}$, $M_{BH} - L_{H,bulge}$, and $M_{BH} - M_{*}$, using host galaxy parameters from the literature.  We find that though individual galaxies often fall within the scatter, as a population they are significantly offset towards higher black hole masses than isolated systems.  Several systems have central masses that sit considerably above scaling relations.  

Under the assumption that these galaxies follow black hole scaling relations before and after their mergers, this suggests that black holes grow more quickly than their host galaxies during a major gas-rich merger.  This does not line up with theoretical reasoning that star formation has easier access to inflowing gas \citep[e.g.][]{Cen12,Hopkins12} or that a final burst of black hole growth is likely to shut off star formation \citep[``quasar-mode feedback";][]{Hopkins05,Springel05}, unless that feedback is delayed due to viscosity in the accretion disk \citep{Power11,Wurster13} or similar.  We conclude that black hole fueling begins early in a gas-rich merger, and can initially outpace any simultaneous starbursting or bulge formation.

These results have focused on a sample of gas-rich mergers in late stages of merging.  To trace black hole growth through the entire merger sequence, a larger sample of black hole masses should be obtained for systems at a range of merger stages.  Such an observational sample could provide the basis for distinguishing between prescriptions for black hole accretion and feedback physics in galaxy merger simulations \citep[e.g.][]{Wurster13_comparison}. 
  
  
\acknowledgements
A.M.M. acknowledges useful conversations with George Privon, Sebastian Haan, Tiantian Yuan, Stephanie Juneau, Kevin Schawinski, David Rosario, and Simona Gallerani.  We enthusiastically thank the staff of the W. M. Keck Observatory and its AO team, for their dedication and hard work.  Data presented herein were obtained at the W. M. Keck Observatory, which is operated as a scientific partnership among the California Institute of Technology, the University of California, and the National Aeronautics and Space Administration.  The Observatory and the Keck II Laser Guide Star AO system were both made possible by the generous financial support of the W. M. Keck Foundation.  The authors wish to extend special thanks to those of Hawai'ian ancestry on whose sacred mountain we are privileged to be guests.  Without their generous hospitality, the observations presented herein would not have been possible.  This work was supported in part by the National Science Foundation Science and Technology Center for Adaptive Optics, managed by the University of California at Santa Cruz under cooperative agreement AST 98-76783.  This material is based in part upon work supported by the National Science Foundation under award number AST-0908796.

{\it Facility:} \facility{Keck:I,II (Laser Guide Star Adaptive Optics, OSIRIS)}

\appendix
\section{Dynamical Mass Models}
\label{BHmodelfigs}

\subsection{Fits and Residuals}
\begin{figure}[ht]
\centering
\includegraphics[scale=0.8]{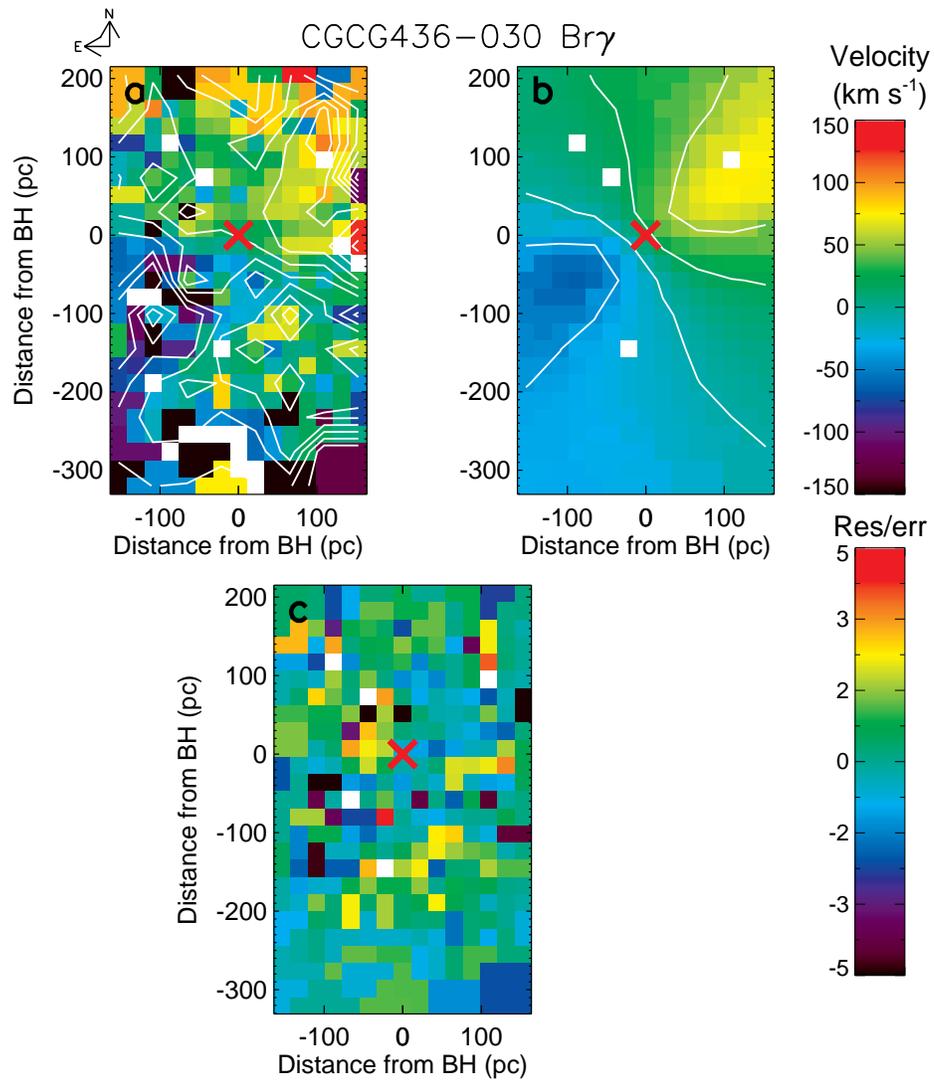}
\caption[A1: Disk Modeling of CGCG436-030's Gas]{a) Map of \brg~velocities of the inner region of CGCG436-030. b) Map of model velocities of best fit black hole model. c) Residual map of best-fit model divided by errors in velocity.  In panels a) and b), velocity contours are marked in white.  In each panel, a red X marks the position of the black hole.  
}
\label{model1}
\end{figure}

In this section, we show the best-fit dynamical mass models for each galaxy for each modeling technique in Figures~\ref{model1}-\ref{model2}.  As discussed in Section~\ref{results}, each black hole was dynamically fit using a subset of the following methods: thin disk fitting to \brg, \paa, or \feii~velocity map, thin disk fitting to \molhy~velocity map, thin disk fitting to stellar velocity map, JAM modeling to stellar velocity and velocity dispersion maps.  Galaxies with multiple tracers show reasonable agreement in measured masses, which can be seen in Table~\ref{tbl:mbh} and Figures~\ref{Msigma}--\ref{Mtotalmass}.  Residuals from each fit, divided by the measurement error, are included in these figures and demonstrate the goodness-of-fit of each model.  The fitted disk parameters are also given in Table~\ref{tbl:disks}.

We note that many galaxies have complex kinematic structures that deviate from pure rotation in one or more of the tracers.  These can be due to tidal streams or spiral arms along the line-of-sight, or to inflow or outflow when seen in gas tracers.  Where possible, we have masked these out to avoid biasing the disk models.  Masked spaxels appear white in the figures.  A full analysis of the deviations from pure rotation and subsequent implications for inflow or outflow will be presented in a future paper.

The positions of the black holes, plotted at (0,0) in each of the following figures, were determined primarily from the kinematic centers rather than the photometric centers.  However, in the case of IRASF17207-0014, the kinematics are sufficiently complex that the two black holes were fixed to the photometric center of each disk measured with GALFIT in \citet{nucleardisks}.

\begin{figure}[ht]
\centering
\includegraphics[scale=0.8]{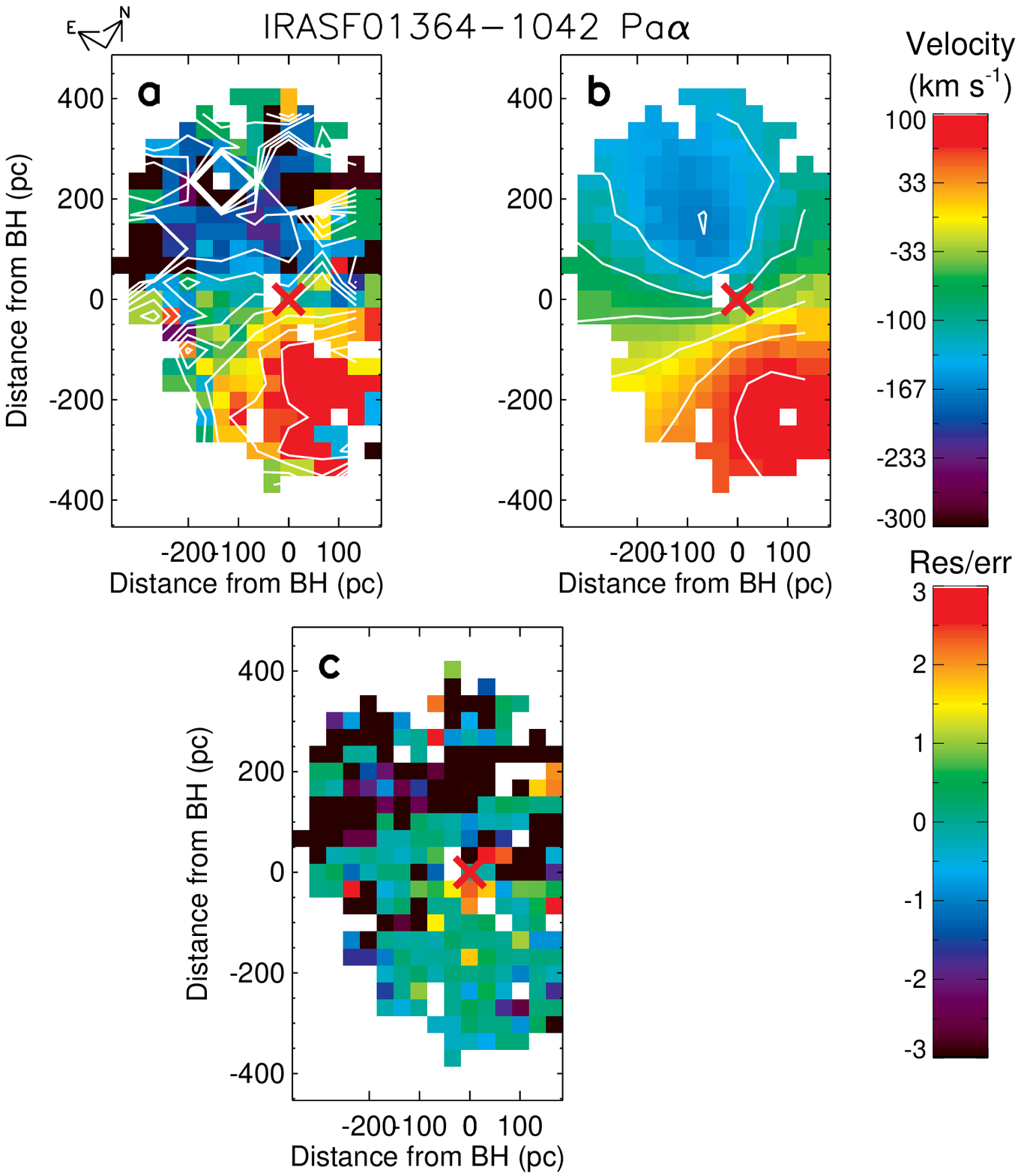}
\caption[A1: Disk Modeling of IR01364's Gas]{a) Map of \paa~velocities of the inner region of IRASF01364-1042. b) Map of model velocities of best fit black hole model. c) Residual map of best-fit model divided by errors in velocity.  In panels a) and b), velocity contours are marked in white.  In each panel, a red X marks the position of the black hole.  
}
\end{figure}

\begin{figure}[ht]
\centering
\includegraphics[scale=0.8]{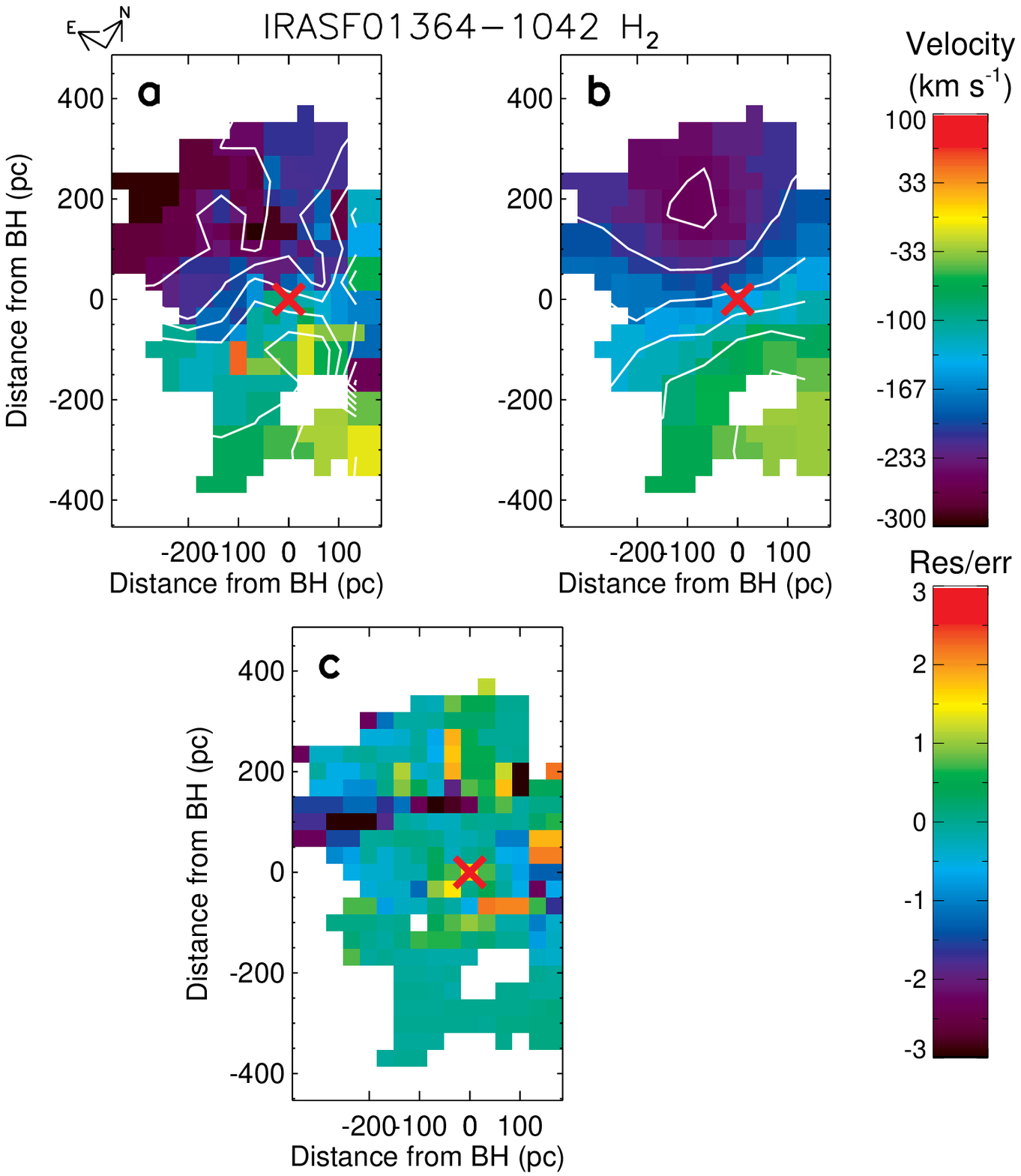}
\caption[A1: Disk Modeling of IR01364's Gas]{a) Map of \molhy~velocities of the inner region of IRASF01364-1042. b) Map of model velocities of best fit black hole model. c) Residual map of best-fit model divided by errors in velocity.  In panels a) and b), velocity contours are marked in white.  In each panel, a red X marks the position of the black hole.  
}
\end{figure}

\begin{figure}[ht]
\centering
\includegraphics[scale=0.8]{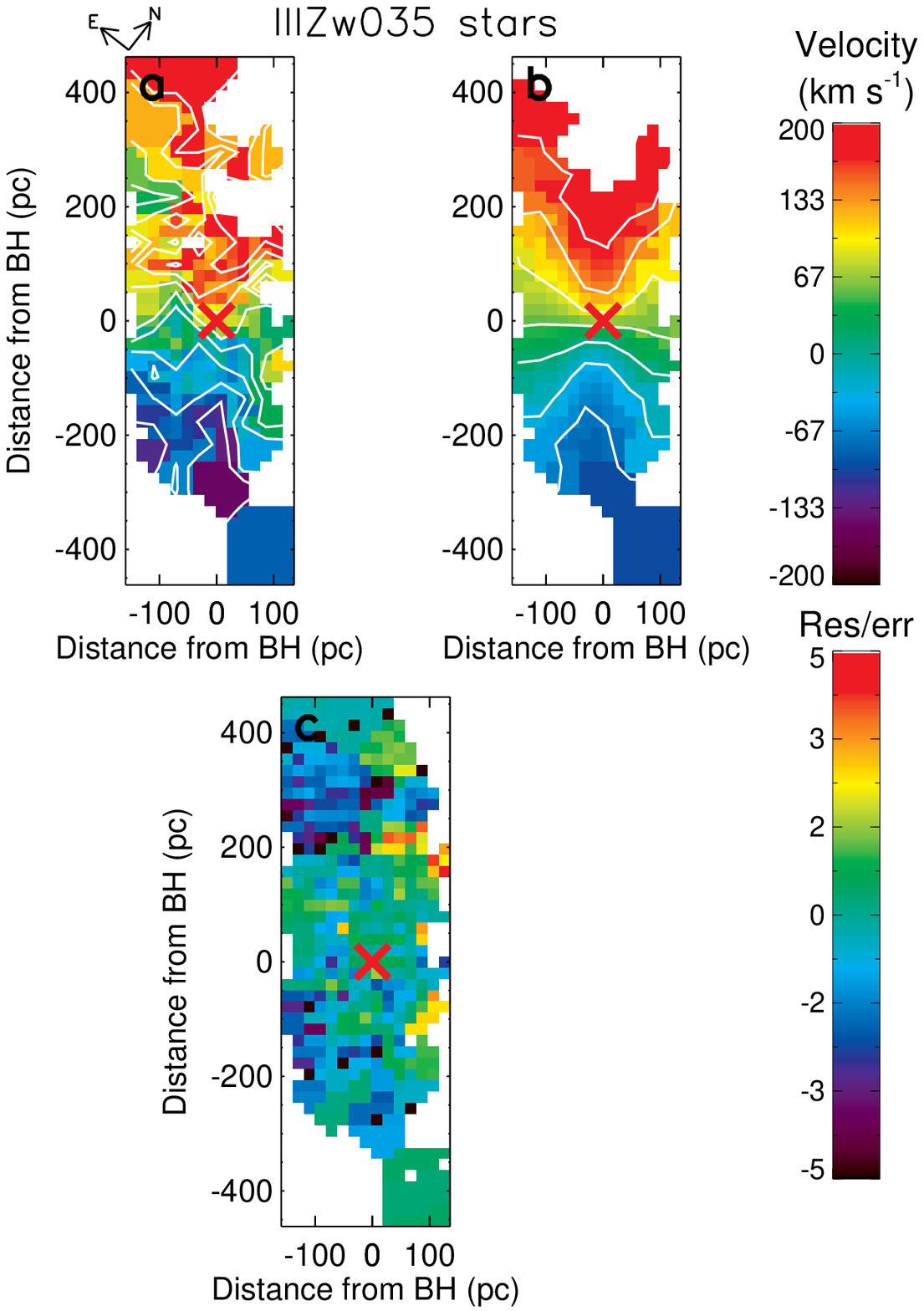}
\caption[A1: Disk Modeling of IIIZw035's Stars]{a) Map of stellar velocities of the inner region of IIIZw035. b) Map of model stellar velocities of best fit black hole model. c) Residual map of best-fit model divided by errors in velocity.  In panels a) and b), velocity contours are marked in white.  In each panel, a red X marks the position of the black hole.  
}
\end{figure}

\begin{figure}[ht]
\centering
\includegraphics[scale=0.75]{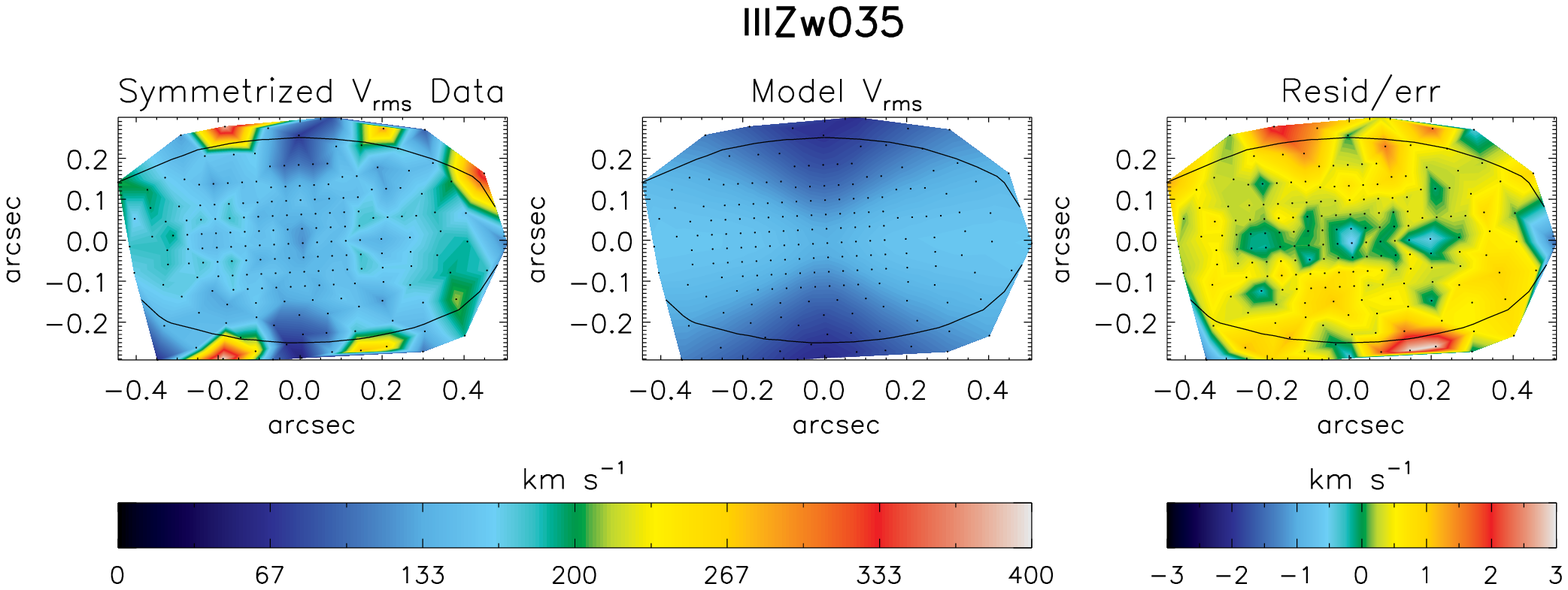}
\caption[A2: JAM Modeling of IIIZw035's Stars]{Left: Map of the symmetrized $v_{rms} = (v^2 + \sigma^2)^{0.5}$ of IIIZw035. Center: Model $v_{rms}$ of galaxy with black hole. Right: Residual map of model - data divided by errors in $v_{rms}$.  All three panels are centered on the black hole and have been rotated so the major axis is horizontal.
}
\end{figure}
\clearpage

\begin{figure}[ht]
\centering
\includegraphics[scale=0.70]{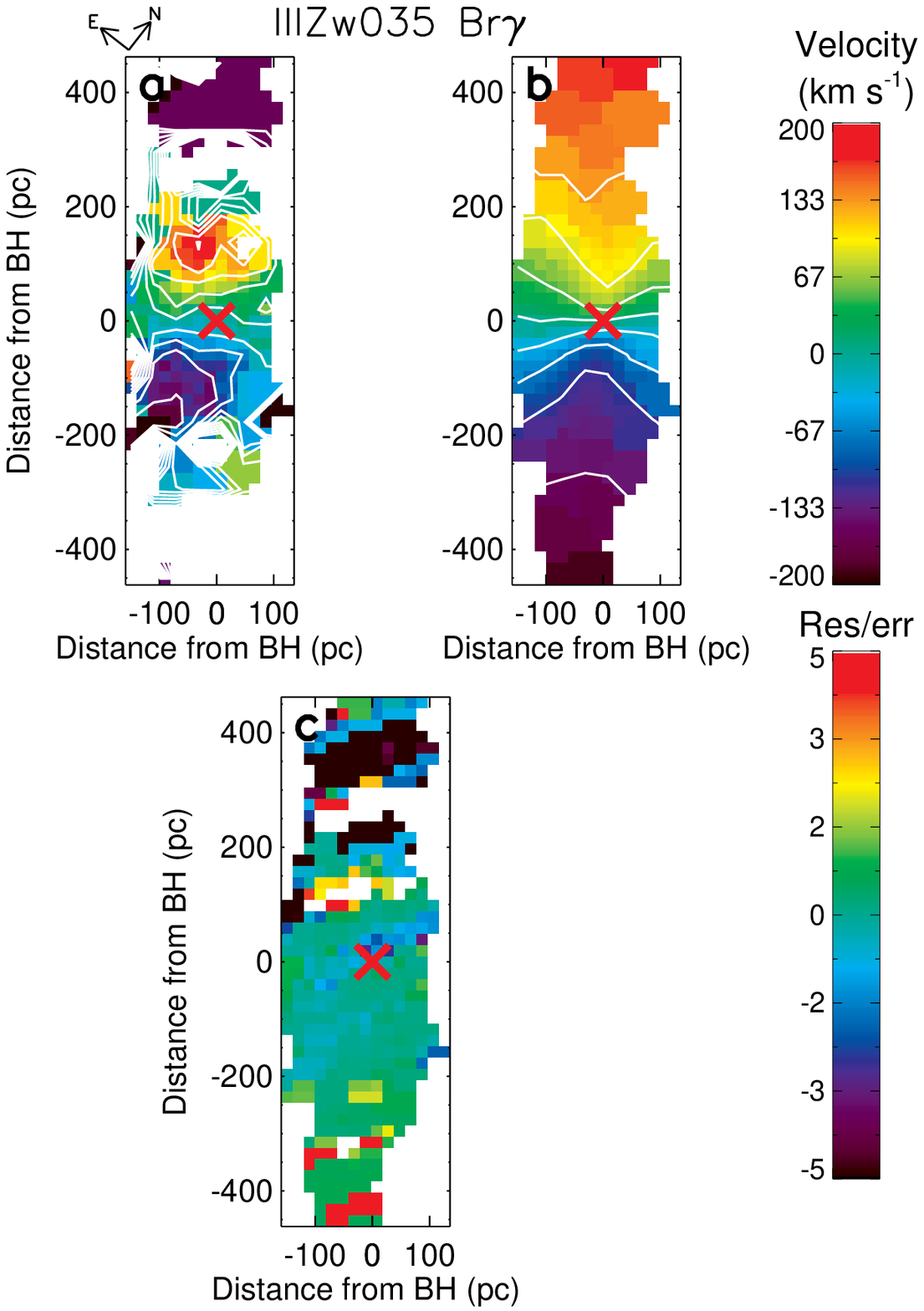}
\caption[A3: Disk Modeling of IIIZw035's \brg]{a) Map of \brg~velocities of the inner region of IIIZw035. b) Map of model velocities of best fit black hole model. c) Residual map of best-fit model divided by errors in velocity.  In panels a) and b), velocity contours are marked in white.  In each panel, a red X marks the position of the black hole.  
}
\end{figure}

\begin{figure}[ht]
\centering
\includegraphics[scale=0.70]{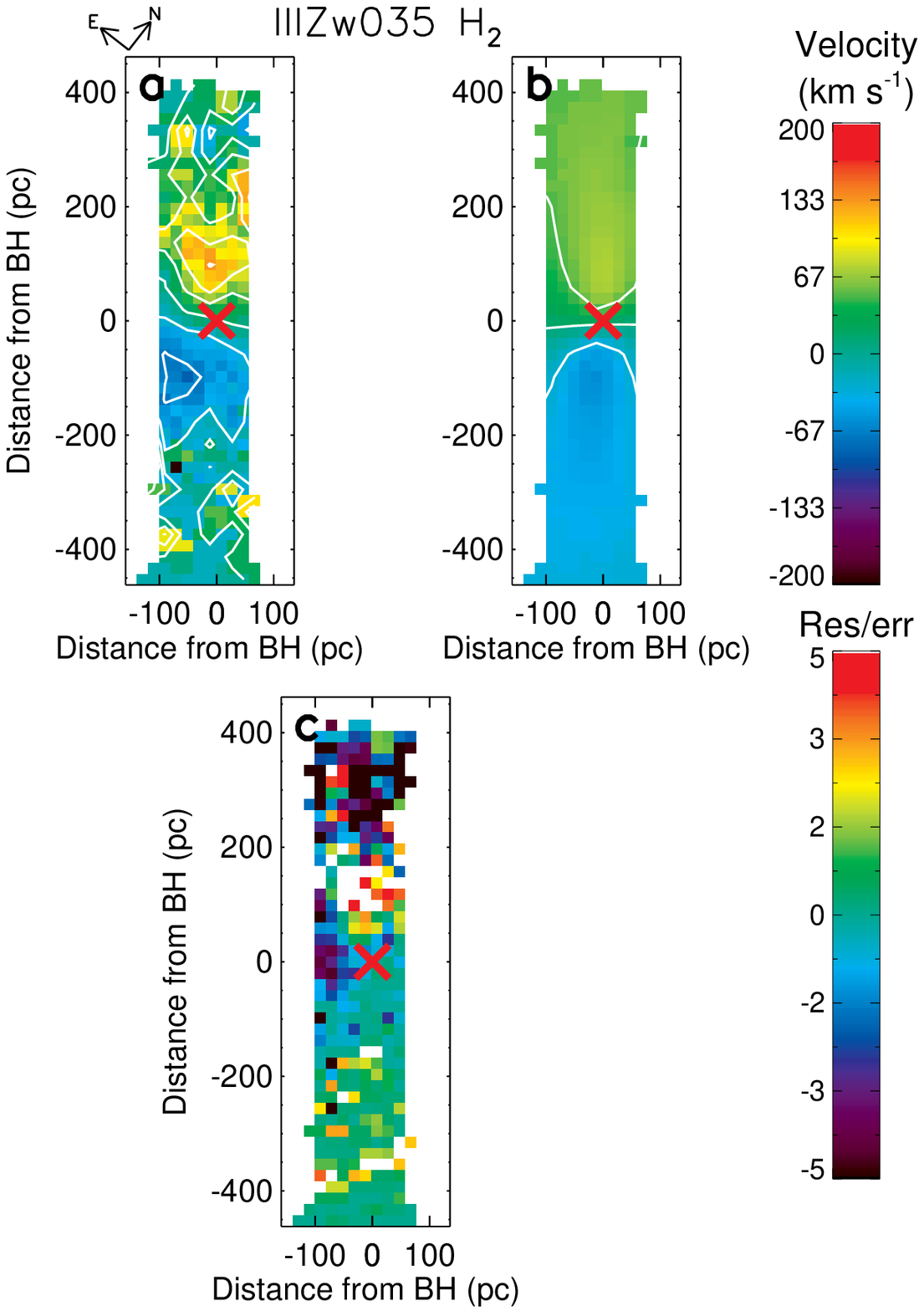}
\caption[A3: Disk Modeling of IIIZw035's \molhy]{a) Map of \molhy~velocities of the inner region of IIIZw035. b) Map of model velocities of best fit black hole model. c) Residual map of best-fit model divided by errors in velocity.  In panels a) and b), velocity contours are marked in white.  In each panel, a red X marks the position of the black hole.  
}
\end{figure}

\begin{figure}[ht]
\centering
\includegraphics[scale=0.8]{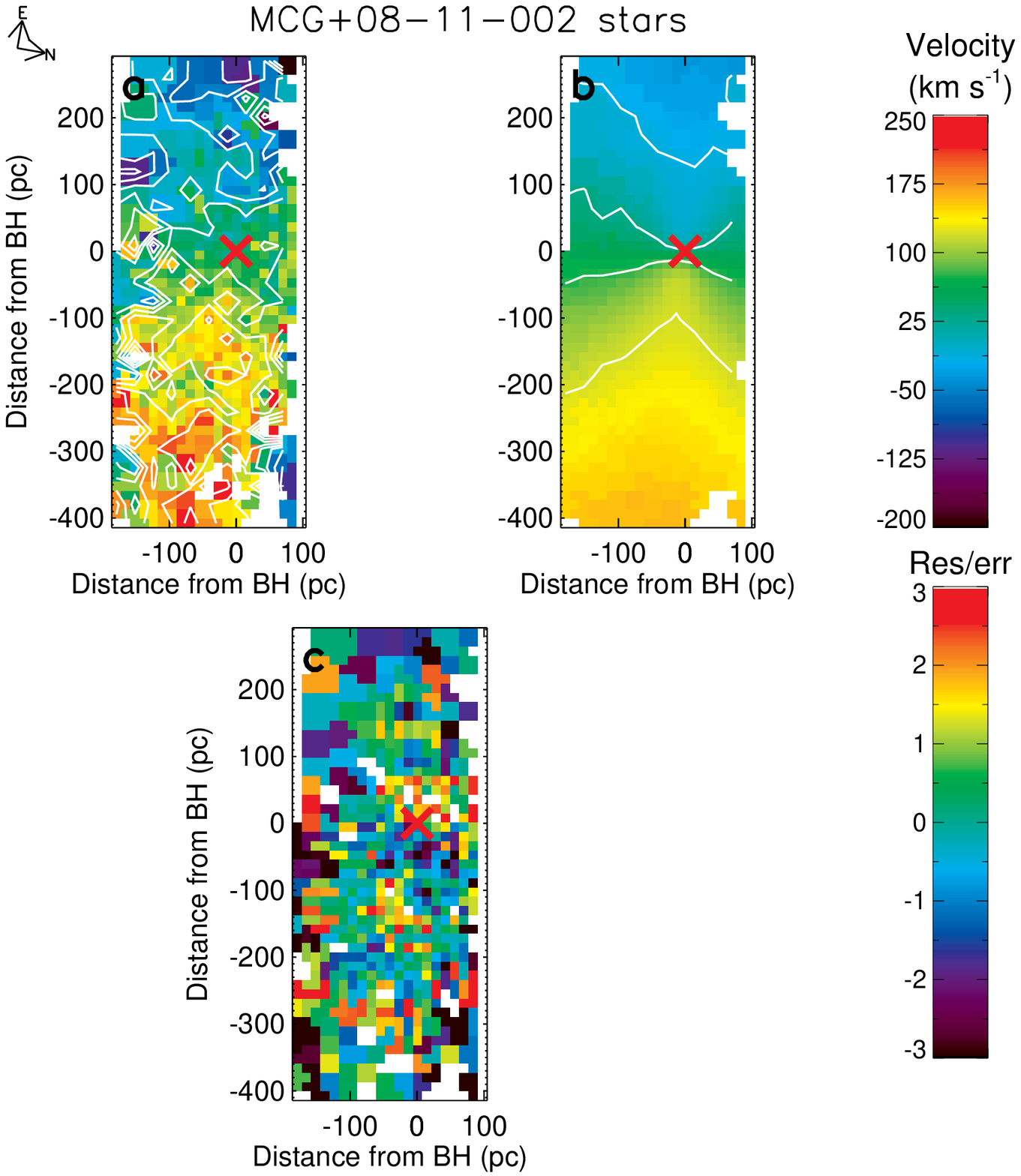}
\caption[A1: Disk Modeling of MCG+08-11-002's Stars]{a) Map of stellar velocities of the inner region of MCG+08-11-002. b) Map of model velocities of best fit black hole model. c) Residual map of best-fit model divided by errors in velocity.  In panels a) and b), velocity contours are marked in white.  In each panel, a red X marks the position of the black hole.  
}
\end{figure}

\begin{figure}[ht]
\centering
\includegraphics[scale=0.75]{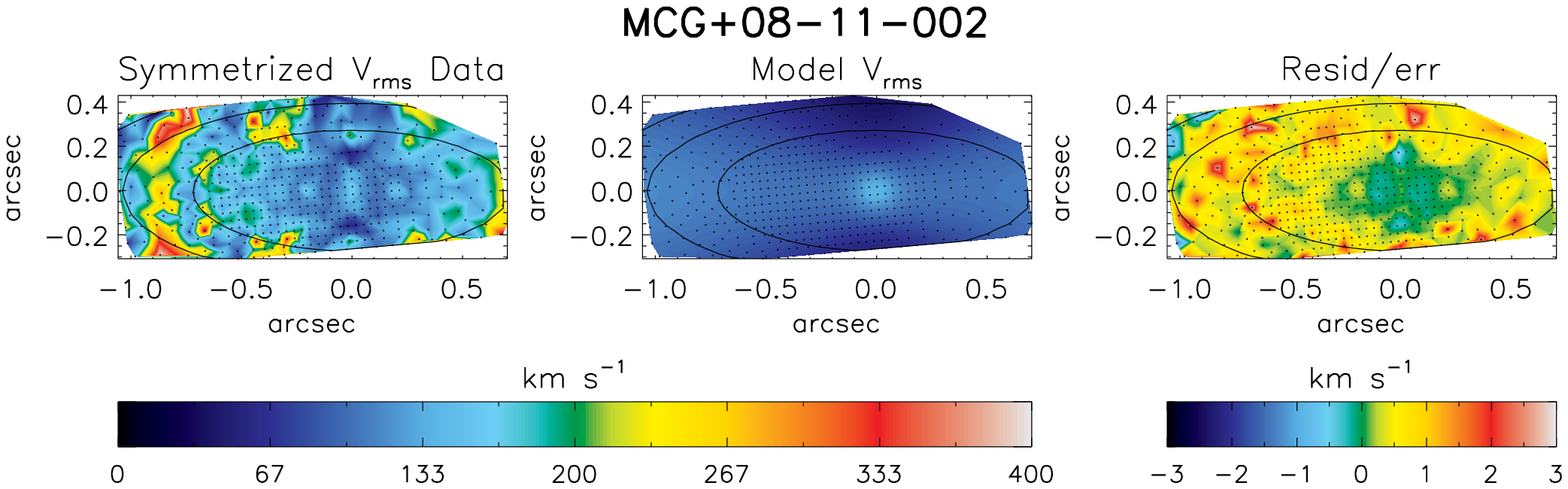}
\caption[A2: JAM Modeling of MCG+08-11-002's Stars]{Left: Map of the symmetrized $v_{rms} = (v^2 + \sigma^2)^{0.5}$ of MCG+08-11-002. Center: Model $v_{rms}$ of galaxy with black hole. Right: Residual map of model - data divided by errors in $v_{rms}$.  All three panels are centered on the black hole and have been rotated so the major axis is horizontal.
}
\end{figure}

\begin{figure}[ht]
\centering
\includegraphics[scale=0.8]{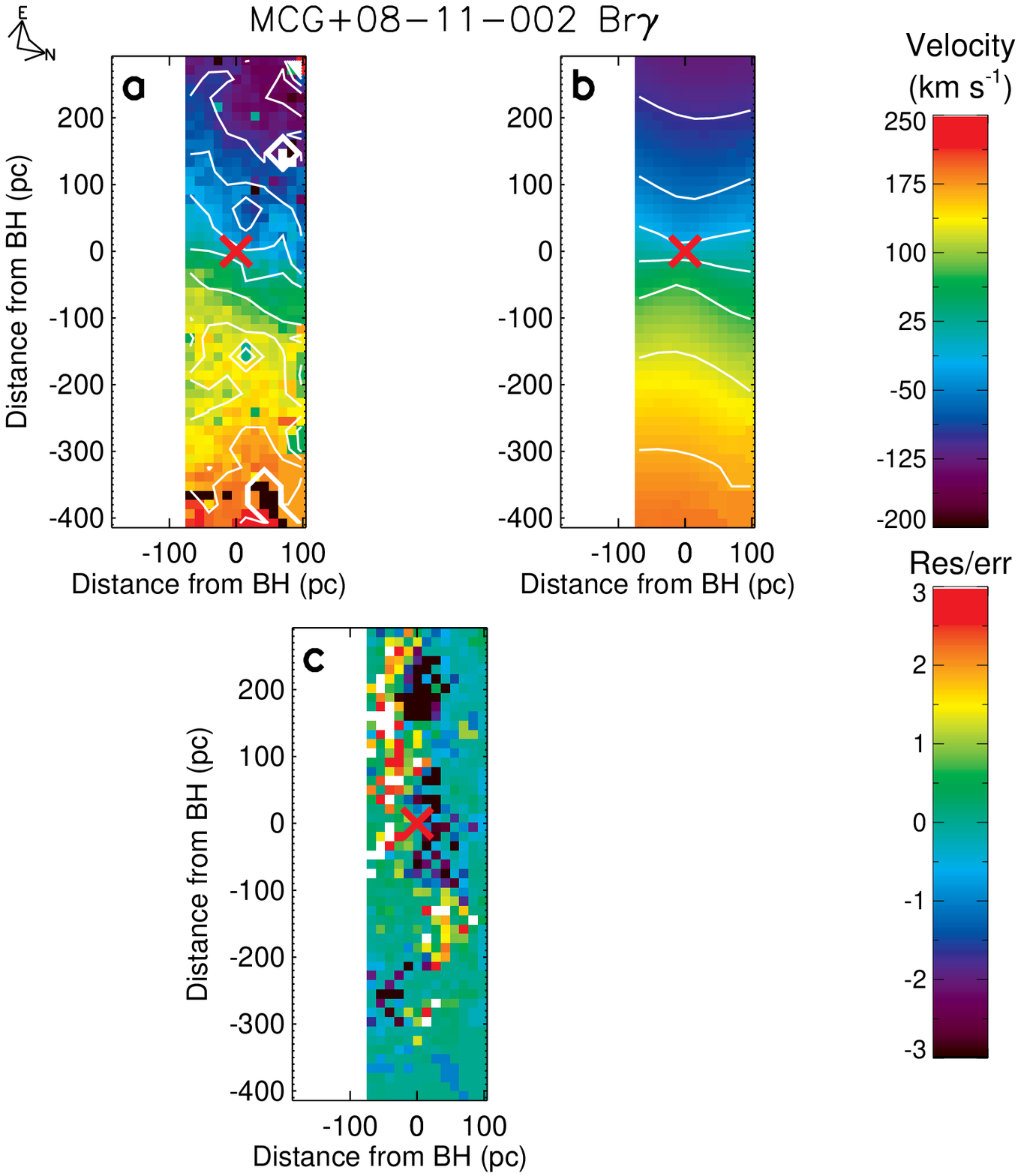}
\caption[A1: Disk Modeling of MCG+08-11-002's Gas]{a) Map of \brg~velocities of the inner region of MCG+08-11-002. b) Map of model velocities of best fit black hole model. c) Residual map of best-fit model divided by errors in velocity.  In panels a) and b), velocity contours are marked in white.  In each panel, a red X marks the position of the black hole.  
}
\end{figure}

\begin{figure}[ht]
\centering
\includegraphics[scale=0.8]{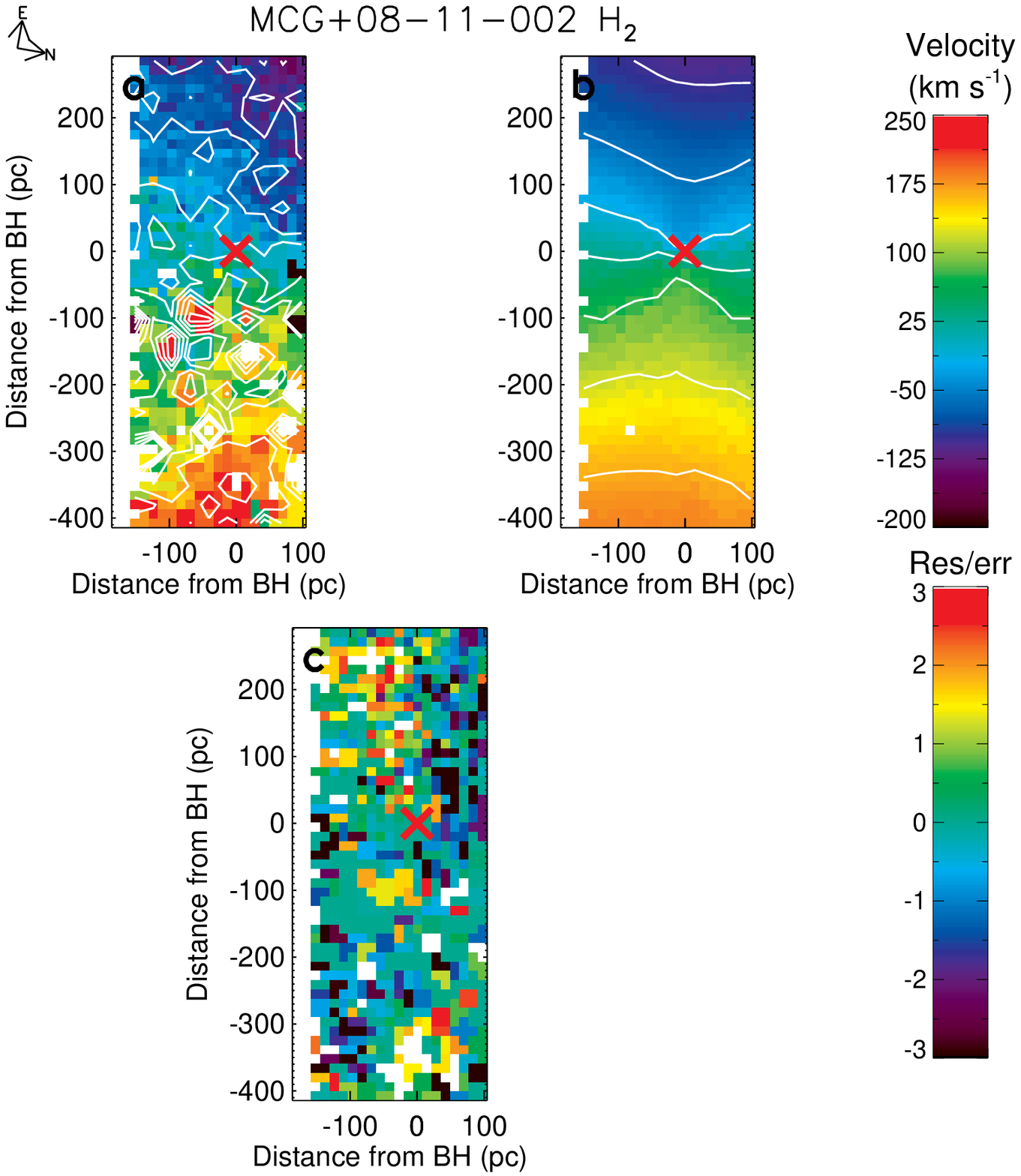}
\caption[A1: Disk Modeling of MCG+08-11-002's Gas]{a) Map of \molhy~velocities of the inner region of MCG+08-11-002. b) Map of model velocities of best fit black hole model. c) Residual map of best-fit model divided by errors in velocity.  In panels a) and b), velocity contours are marked in white.  In each panel, a red X marks the position of the black hole.  
}
\end{figure}

\begin{figure}[ht]
\centering
\includegraphics[scale=0.75]{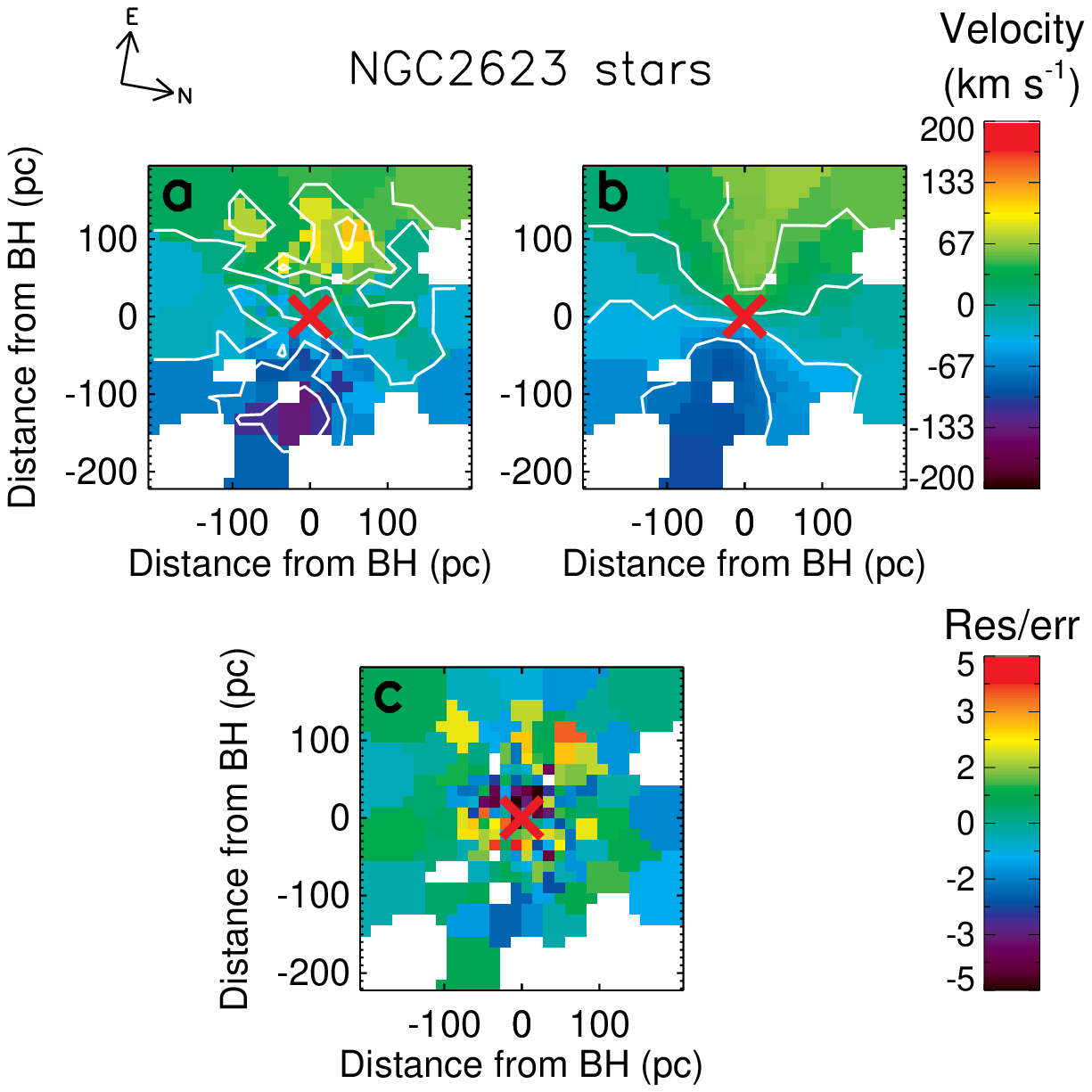}
\caption[B1: Disk Modeling of NGC~2623's Stars]{a) Map of stellar velocities of the inner region of NGC2623. b) Map of model stellar velocities of best fit black hole model. c) Residual map of best-fit model divided by errors in velocity.  In panels a) and b), velocity contours are marked in white.  In each panel, a red X marks the position of the black hole. }
\end{figure}


\begin{figure}[ht]
\centering
\includegraphics[scale=0.65]{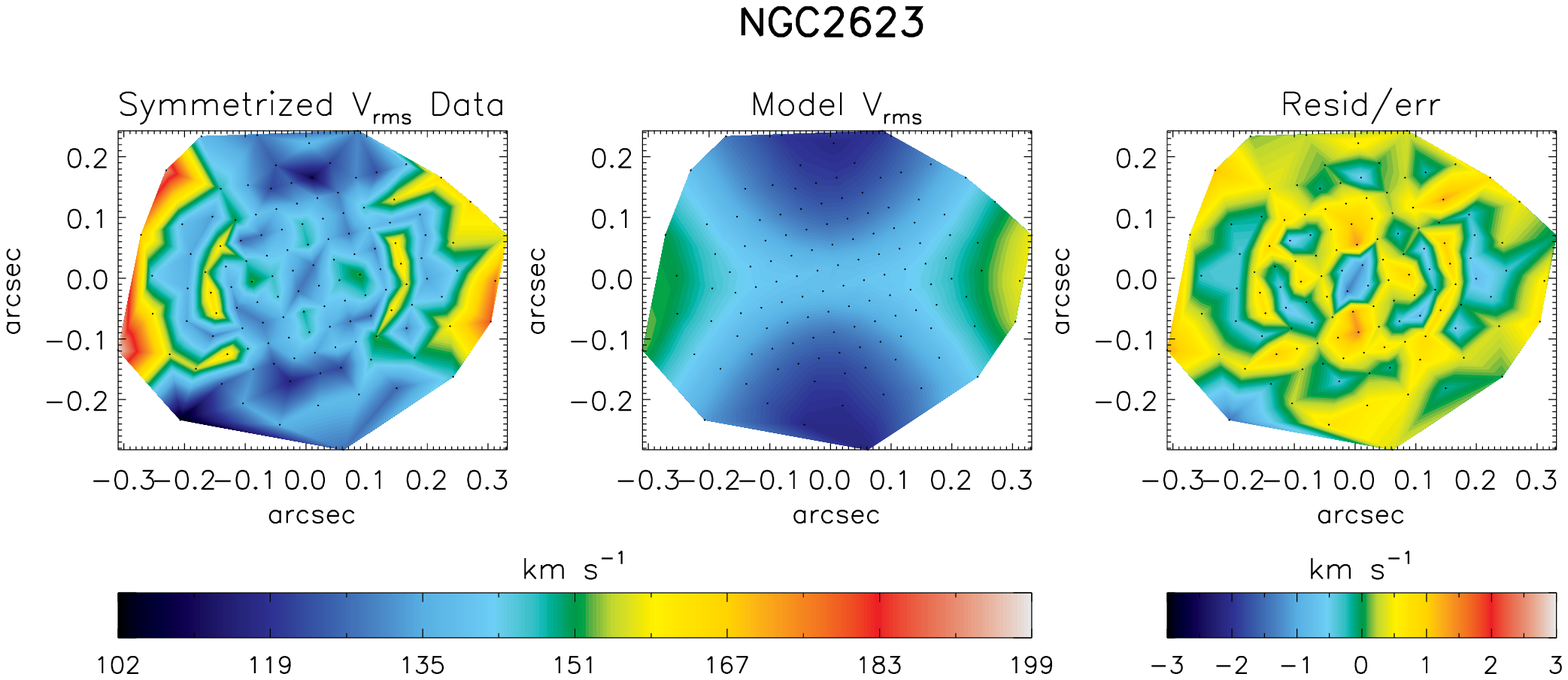}
\caption[b2: JAM Modeling of NGC2623's Stars]{Left: Map of the symmetrized $v_{rms} = (v^2 + \sigma^2)^{0.5}$ of NGC2623. Center: Model $v_{rms}$ of galaxy with black hole. Right: Residual map of model - data divided by errors in $v_{rms}$.  All three panels are centered on the black hole and have been rotated so the major axis is horizontal.
}
\end{figure}

\begin{figure}[ht]
\centering
\includegraphics[scale=0.75]{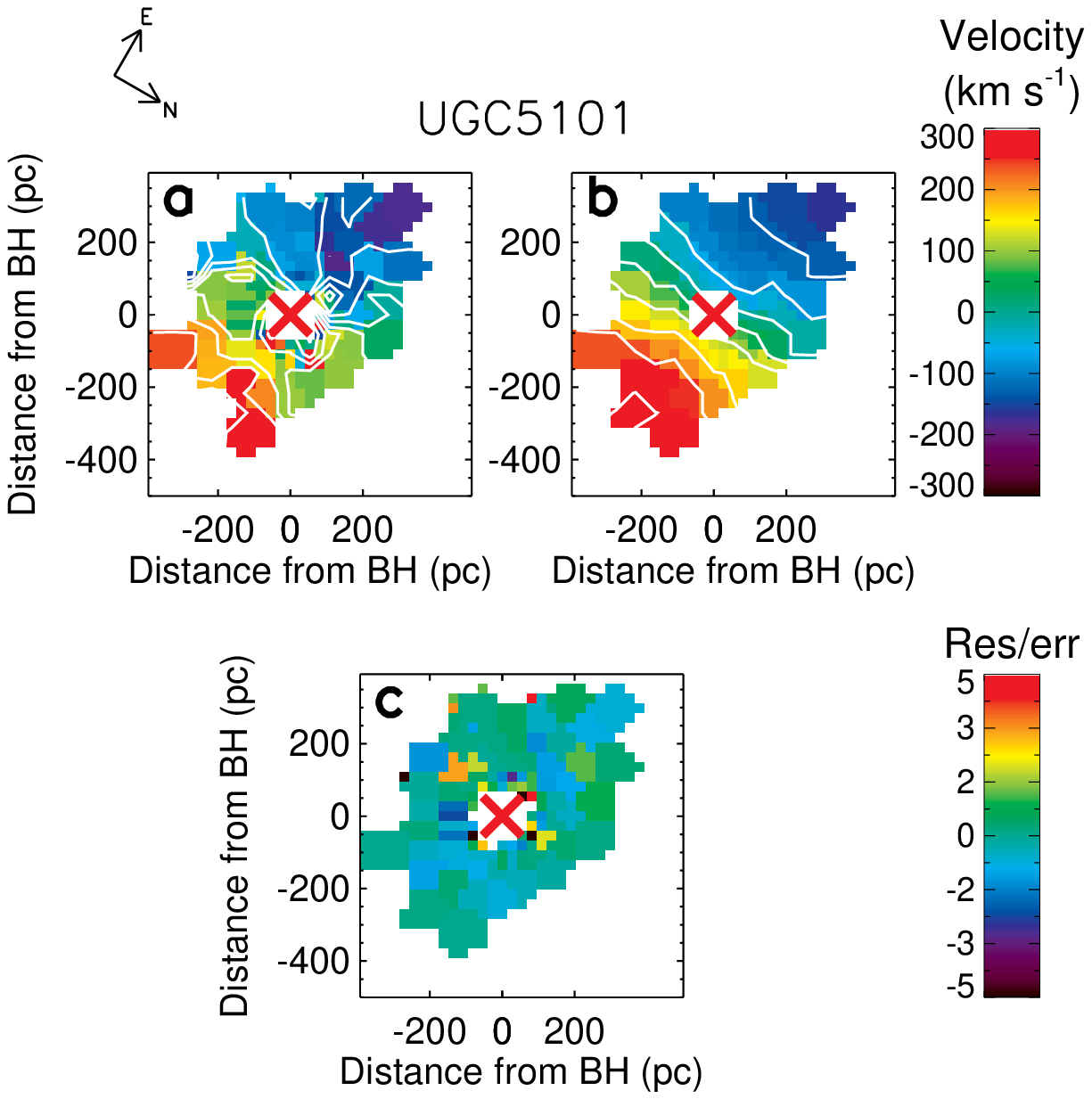}
\caption[B1: Disk Modeling of UGC5101's Stars]{a) Map of stellar velocities of the inner region of UGC5101. b) Map of model stellar velocities of best fit black hole model. c) Residual map of best-fit model divided by errors in velocity.  In panels a) and b), velocity contours are marked in white.  In each panel, a red X marks the position of the black hole. }

\end{figure}

\begin{figure}[ht]
\centering
\includegraphics[scale=0.65]{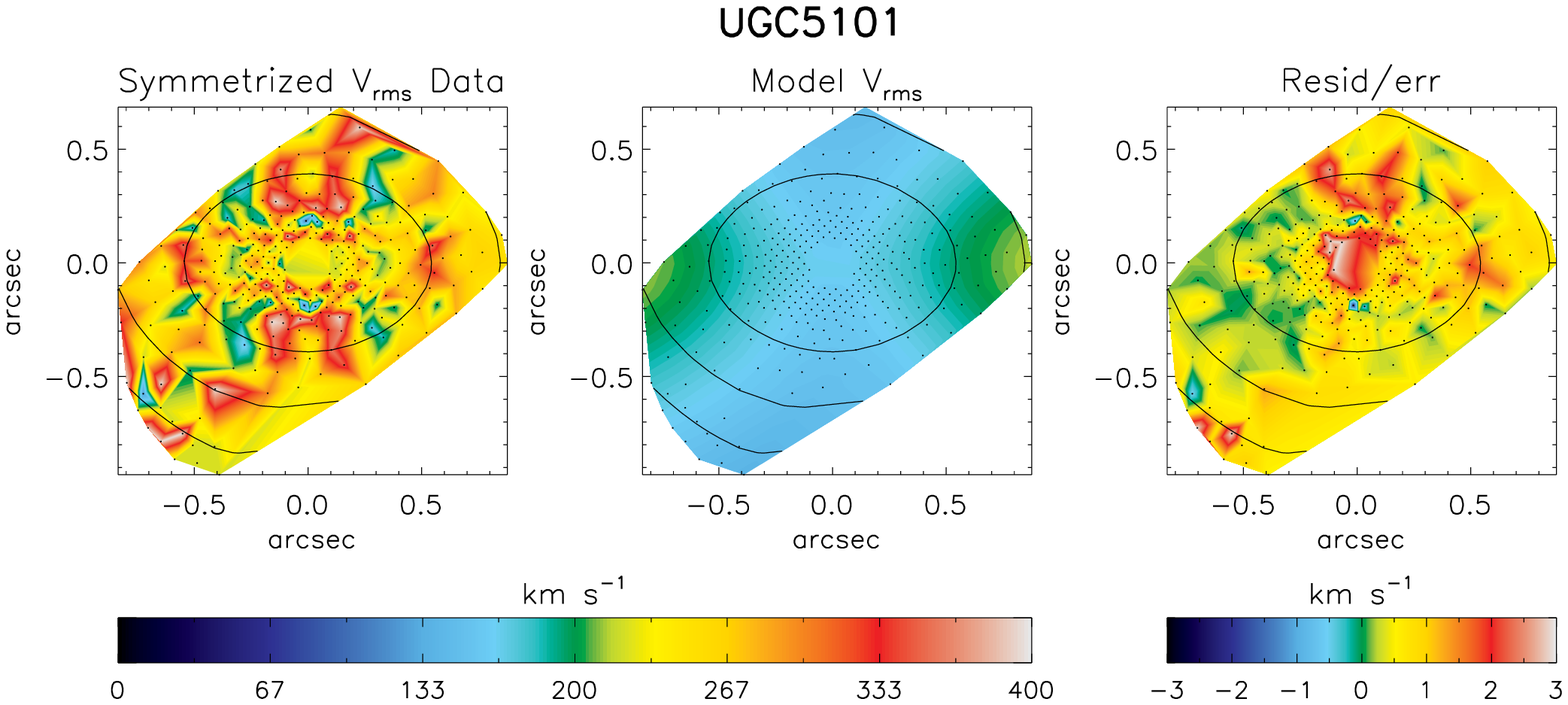}
\caption[b2: JAM Modeling of UGC5101's Stars]{Left: Map of the symmetrized $v_{rms} = (v^2 + \sigma^2)^{0.5}$ of UGC5101. Center: Model $v_{rms}$ of galaxy with black hole. Right: Residual map of model - data divided by errors in $v_{rms}$.  All three panels are centered on the black hole and have been rotated so the major axis is horizontal.
}
\end{figure}

\begin{figure}[ht]
\centering
\includegraphics[scale=0.75]{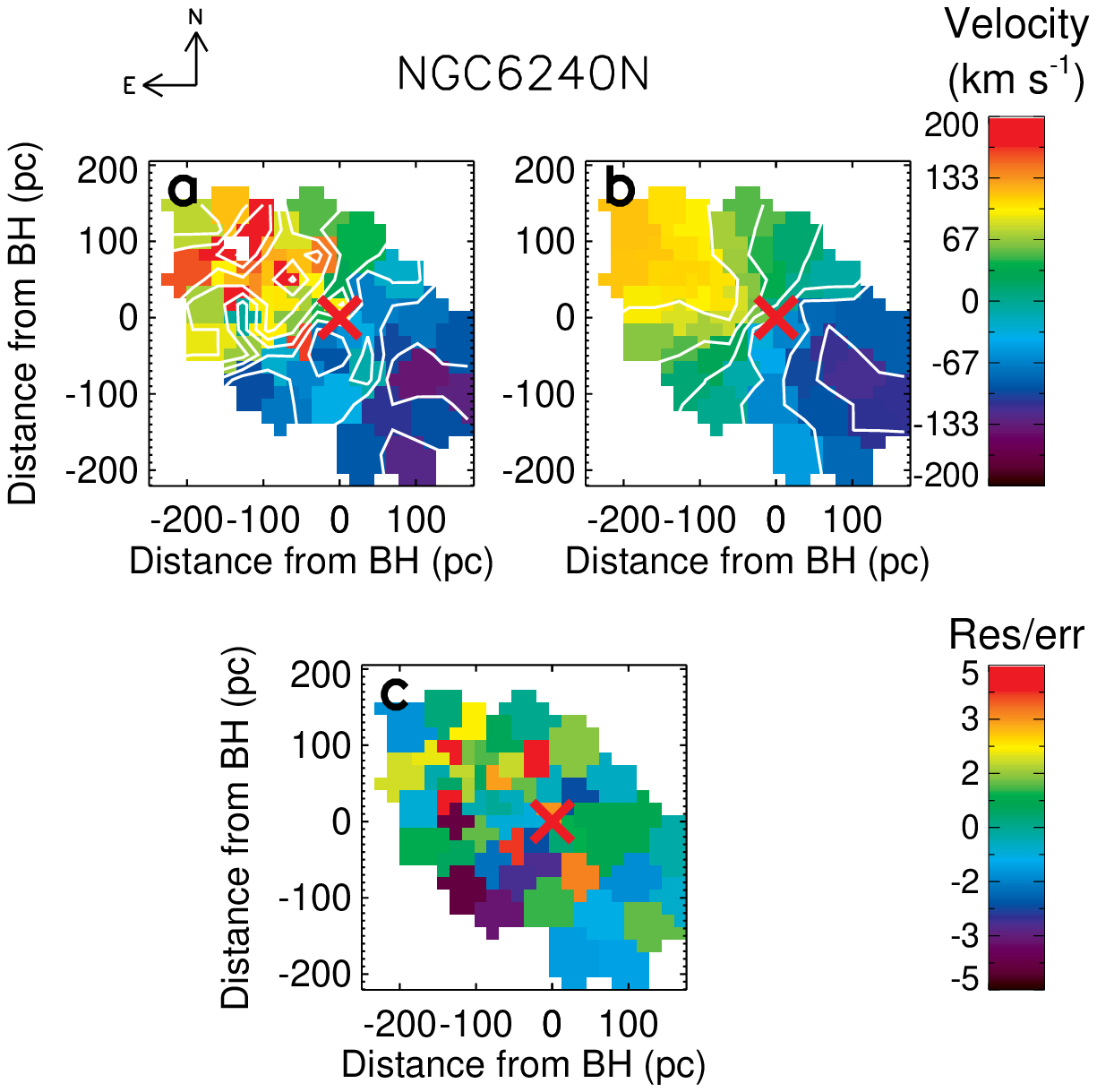}
\caption[C1: Disk Modeling of NGC~6240 N's Stars]{a) Map of stellar velocities of the inner region of NGC~6240N. b) Map of model stellar velocities of best fit black hole model. c) Residual map of best-fit model divided by velocity errors.  In panels a) and b), velocity contours are marked in white.  In each panel, a red X marks the position of the black hole. }
\end{figure}

\begin{figure}[ht]
\centering
\includegraphics[scale=0.70]{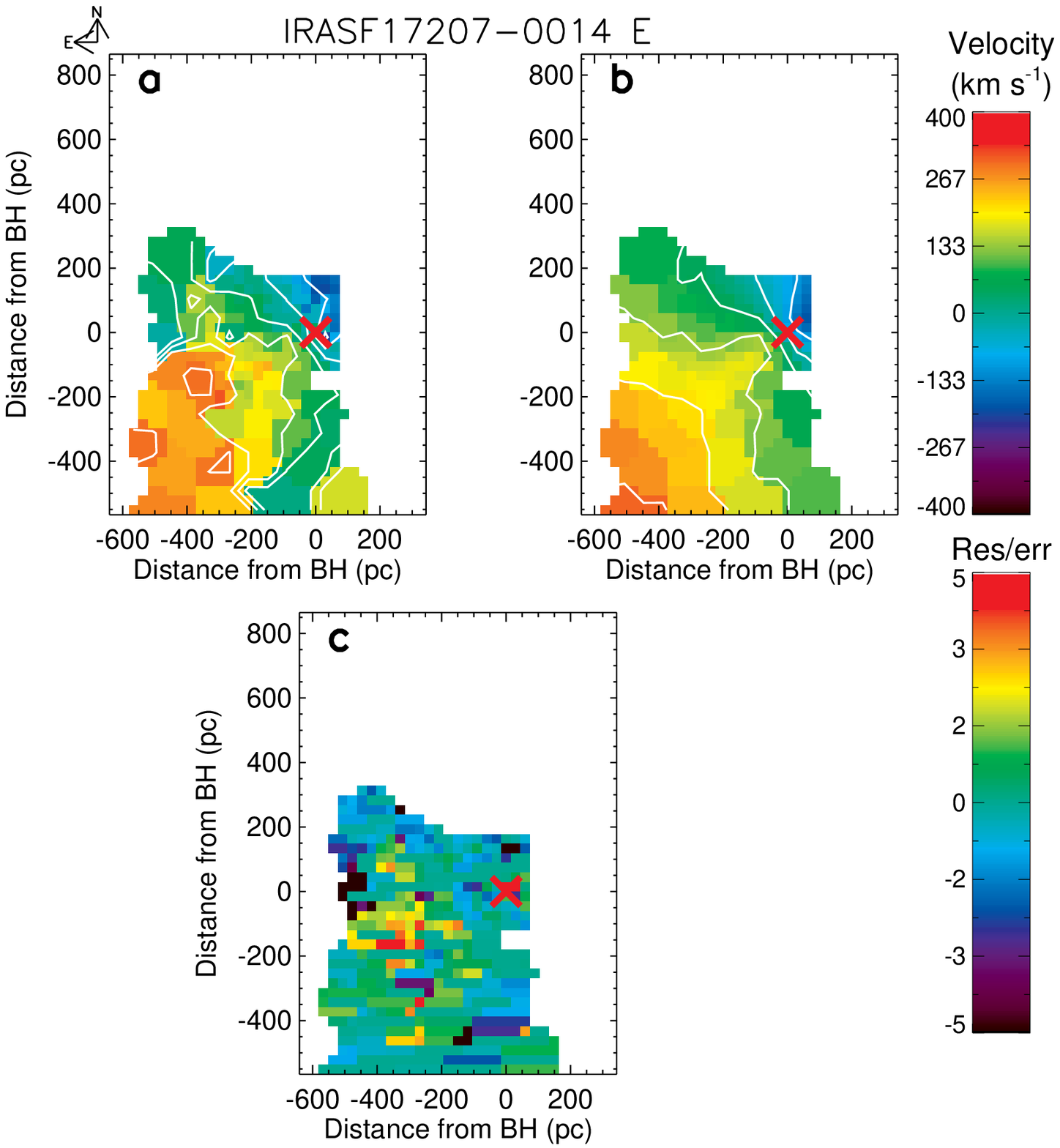}
\caption[A3: Disk Modeling of IR17207's \feii]{a) Map of \feii~velocities of the inner region of IRAS F17207-0014. b) Map of model velocities of best fit black hole model. c) Residual map of best-fit model divided by errors in velocity.  In panels a) and b), velocity contours are marked in white.  In each panel, a red X marks the position of the black hole.  We have masked out regions in order to fit only the eastern black hole.
}
\label{model2}
\end{figure}

\begin{figure}[ht]
\centering
\includegraphics[scale=0.70]{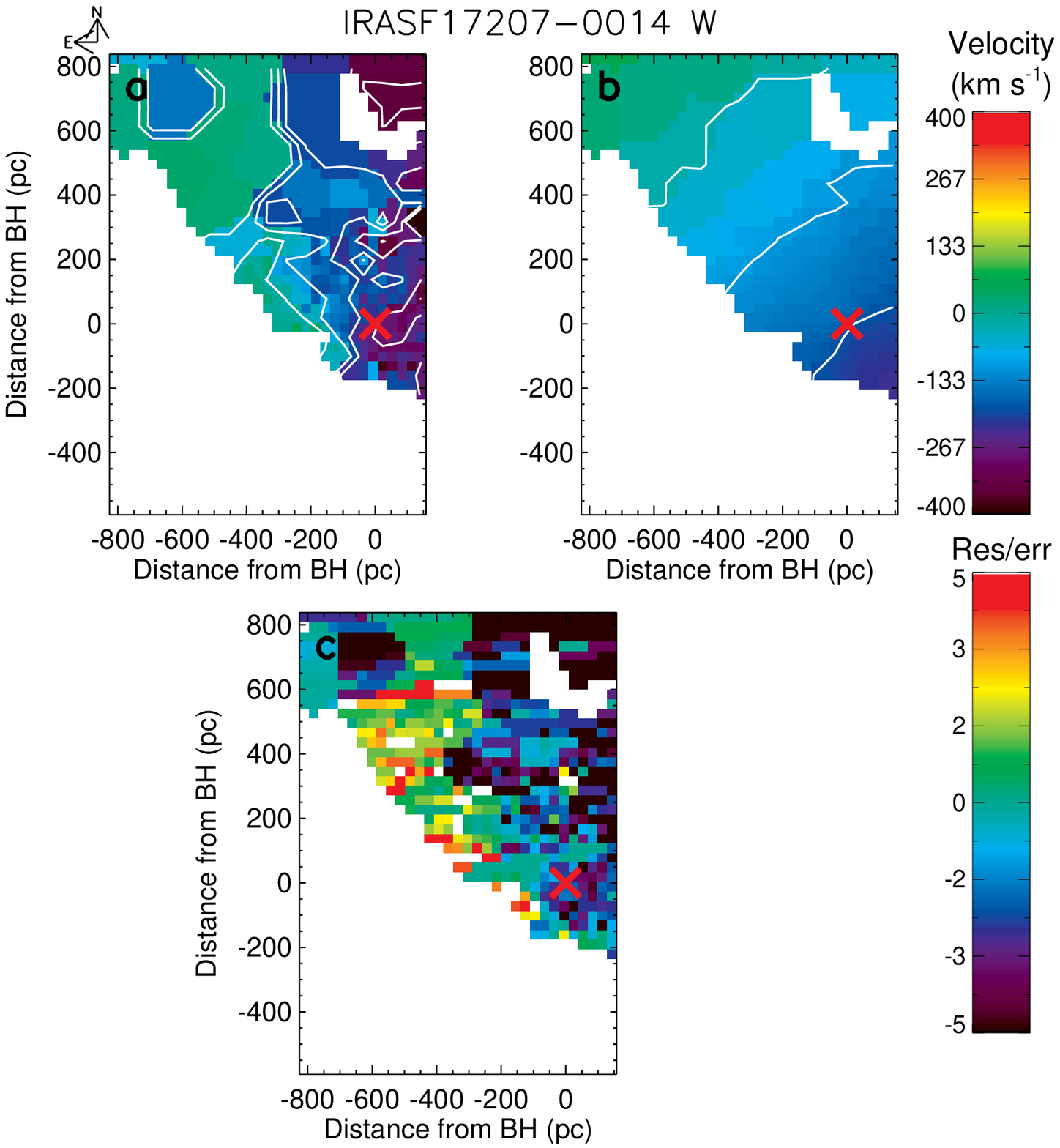}
\caption[A3: Disk Modeling of IR17207's \feii]{a) Map of \feii~velocities of the inner region of IRAS F17207-0014. b) Map of model velocities of best fit black hole model. c) Residual map of best-fit model divided by errors in velocity.  In panels a) and b), velocity contours are marked in white.  In each panel, a red X marks the position of the black hole.  We have masked out regions in order to fit only the western black hole.
}
\label{model2}
\end{figure}

\subsection{Black Hole Spheres of Influence}

One key feature of our thin disk black hole mass models is the inclusion of a radially-varying mass component to incorporate the mass profile of the nuclear disk.  (Our JAM models also incorporate galactic mass, but by fitting a mass-to-light ratio.)  We note that our fits use only a single power-law fit to the mass profile, so any pile-up of matter below our resolution limit will be considered part of the point mass.  We include the mass profiles measured here for completeness, and for the purposes of calculating the implied radius of the sphere of influence of each black hole.  Recall that the mass profile is fit with the form $M(r) = \rho_{0} r^{-\gamma}$.  In Table~\ref{tbl:disks} we report the normalization and power law fit for each of these parameters.  For clarity, instead of reporting $\rho_{0}$, we report the mass in the disk enclosed at 100 parsecs.

A black hole's sphere of influence is defined as the region within which the black hole contributes 50\% of the mass.  We can therefore use our thin disk mass profiles to calculate the radius of the implied sphere of influence.  This is interesting both for the purposes of understanding black hole dynamics and for confirming that we indeed have sufficient resolution to measure the black hole masses appropriately.

As the fitted mass profiles of the disk and black hole masses are not independent, there is an issue of covariance to consider.  That is, when a model fits a higher black hole mass, it will likely put less mass in the disk, and vice versa.  When calculating errors on each of these parameters, we marginalize over the others: our black hole mass errors are calculated from the width of the black hole mass distribution in our Monte Carlo simulation, taking into account all possible fitted disk profiles.  Similarly, we report our errors on disk profile parameters using the widths of the distributions taking into account all fitted black hole masses.  This is an appropriate way to handle covariance when drawing conclusions about the black hole masses, as is done in this paper.  To account for the covariance issue in the measurement of $r_{infl}$, we calculate $r_{infl}$ for every fit in the Monte Carlo simulation and draw our conclusions based on that.

In one case (IRASF01364-1042), the black hole mass is comparable or larger than the entire nuclear disk. The calculated radius of influence is thus artificially large, because presumably the mass profile of the galaxy changes beyond our field of view.  The \molhy~fit to IIIZw035 is also artificially large; we attribute this to streaming motion in the \molhy~emission compromising the larger scale fit.  This and other evidence of outflows will be presented in an upcoming paper.  The affected radii are reported in parentheses in Table~\ref{tbl:disks}.

We see that, for our sample, we clearly resolve the spheres of influence of these black holes.  We note that this method of calculating the size of the sphere of influence is uncommon, as a full mass profile is required to determine where $M(r) = M_{BH}$.  Our capacity to resolve these scales exceeded our expectations because the black holes are more massive than predicted.

\begin{deluxetable}{lccccccc}
    \centering
    \tabletypesize{\scriptsize}
    \tablewidth{0pt}
    \tablecolumns{8}
    \tablecaption{Measured Disk Parameters}
    \tablehead{   
      \colhead{Galaxy Name} &
      \colhead{Tracer} &
      \colhead{$M_{BH}$} &
      \colhead{$M_{gal}(<100 pc)$ }  &
      \colhead{$\gamma$} &
      \colhead{$r_{infl}$} &
      \colhead{$r_{infl}$}  & \\
      \colhead{}&
      \colhead{}&
      \colhead{$(M_{\sun})$}&
      \colhead{$(M_{\sun})$}&
      \colhead{}&
      \colhead{(pc)}&
      \colhead{(pixels)} &
      	}
    \startdata
    CGCG436-030 & \brg & $4.59 \substack{+0.52\\-0.48} \times 10^{8} $& $6.0 \pm3.5 \times10^{6}$ & $0.7\pm0.2$ & $390\pm60$ &  $17.9\pm2.8$\\
    IRASF01364-1042 & \brg & $2.37 \substack{+0.01\\-0.1} \times 10^{9}$& $6.2\pm1.1 \times 10^{4}$ & $2.0\tablenotemark{a}$ & ($2.3 \pm 0.1 \times10^{6} $)\tablenotemark{b}& ($7.0\pm0.3\times10^{4}) $\\
     & \molhy & $2.12 \substack{+0.06\\-0.14} \times 10^{9}$ & $1.0\pm.1 \times10^{5}$& $2.0$ & ($2.0\pm0.1 \times 10^{6}$) & ($6.0\pm0.3\times10^{4}$)  \\
    IIIZw035 & stars - disk & $>6.8\substack{+0.1\\-4.0}\times 10^8 $&$2.7\pm0.4 \times10^{8}$ & 1.0 & $114\pm19$ & $5.9\pm1.0$  \\ 
     &\molhy& $2.59 \substack{+0.03 \\ -0.1} \times 10^{8} $& $1.0\pm0.5 \times10^{5}$ & $1.0$ & ($5060 \pm 70$ ) & ($260\pm3$)\\ 
[0.5ex]      & \brg & $3.4 \substack{+0.4 \\ -0.6} \times10^{8}$ & $1.9\pm0.2 \times10^{8}$ & $1.0$ & $126\pm13$ & $6.4 \pm 0.7$ \\ 
[0.5ex] 
    MCG+08-11-002 & stars - disk & $>8.7 \substack{+3.1\\-3.0} \times 10^{7}$& $ 1.0\pm0.1 \times10^{8}$& $1.0\pm0.1$ & $85\pm19$ & $6.1\pm1.3$ \\
     & \molhy  & $6.9 \substack{+1.0\\-1.6} \times 10^{7}$ &$1.7\pm0.1 \times10^{8}$ &$0.7\pm0.1$ &$61\pm9$ & $4.4\pm0.6$ \\
     & \brg  & $2.3\substack{+1.2\\-0.9} \times 10^7$ & $2.8\pm0.2\times10^{8}$ & $0.9\pm0.1$& $24\pm9$ & $1.7\pm0.6$\\
[0.5ex]     NGC~2623 & stars - disk & $>2.9\substack{+0.1\\-0.7}\times10^8$& $1.4\pm0.2 \times10^{8}$ & $1.0\pm0.1$ & $127\pm14$ & $9.1\pm1.0$  \\
[0.5ex]     UGC5101 & stars - disk & $>6.5 \substack{ +3.5\\-2.1} \times 10^8$& $9.4\pm0.4 \times10^{8} $ & $1.1\pm0.1$ & $80\pm22$ & $3.0\pm0.8$ \\
[0.5ex]     Mrk273 N &\feii & $1.0\pm 0.1\times10^9$ & $1.1\pm0.1\times10^{9}$ &  2.0 & $89\pm11$ & $3.4\pm0.4$ \\
[0.5ex]     NGC~6240 N & stars - disk &$ > 8.8\substack{+0.7 \\-0.1 } \times10^8$  & $1.2\pm0.1 \times10^{8}$ & $0.63\pm0.04$ &  $235\pm8$ & $13.7\pm0.5$  \\
[0.5ex]     NGC~6240 S & stars - disk & $>8.7\pm0.3\times10^8$ & $2.8\pm0.1 \times10^{8}$ & $1.5\pm0.1$ & $212\pm9$ & $12.3\pm0.5$  \\
    IRASF17207-0014 E & \feii & $2.5\substack{+0.03\\-0.3}\times10^9$ & $5.3\pm0.3 \times 10^{7}$ & $0.0\pm0.1$ & $346\pm10$ & $11.6\pm0.3$  \\
    IRASF17207-0014 W & \feii & $8.9\substack{+4.0\\-1.6}\times10^{7}$& $6.1\pm0.4 \times10^{8}$ & $1.1\pm0.1$ &  $142\pm9$ & $4.8\pm0.3$ \\
    \enddata
\tablenotetext{a}{Some fits were run with $\gamma$ fixed to match the light profile; these are thus reported with no error bars.}
\tablenotetext{b}{Radii in parentheses are artificially large because the mass of the black hole is comparable to or larger than the fitted nuclear disk.  Presumably the mass profile of the galaxy changes outside of our field of view, which would reduce the radius of the sphere of influence.  In the case of IIIZw035 \molhy, non-rotational kinematics unseen in other tracers are likely affecting the fit.}
    \label{tbl:disks}
  \end{deluxetable}






\bibliographystyle{apj}

\begin{thebibliography}{}
\expandafter\ifx\csname natexlab\endcsname\relax\def\natexlab#1{#1}\fi

\bibitem[{{Alatalo} {et~al.}(2011){Alatalo}, {Blitz}, {Young}, {Davis},
  {Bureau}, {Lopez}, {Cappellari}, {Scott}, {Shapiro}, {Crocker},
  {Mart{\'{\i}}n}, {Bois}, {Bournaud}, {Davies}, {de Zeeuw}, {Duc}, {Emsellem},
  {Falc{\'o}n-Barroso}, {Khochfar}, {Krajnovi{\'c}}, {Kuntschner}, {Lablanche},
  {McDermid}, {Morganti}, {Naab}, {Oosterloo}, {Sarzi}, {Serra}, \&
  {Weijmans}}]{Alatalo11}
{Alatalo}, K., {Blitz}, L., {Young}, L.~M., {et~al.} 2011, \apj, 735, 88

\bibitem[{{Alexander} {et~al.}(2008){Alexander}, {Brandt}, {Smail}, {Swinbank},
  {Bauer}, {Blain}, {Chapman}, {Coppin}, {Ivison}, \&
  {Men{\'e}ndez-Delmestre}}]{Alexander08}
{Alexander}, D.~M., {Brandt}, W.~N., {Smail}, I., {et~al.} 2008, \aj, 135, 1968

\bibitem[{{Armus} {et~al.}(2009){Armus}, {Mazzarella}, {Evans}, {Surace},
  {Sanders}, {Iwasawa}, {Frayer}, {Howell}, {Chan}, {Petric}, {Vavilkin},
  {Kim}, {Haan}, {Inami}, {Murphy}, {Appleton}, {Barnes}, {Bothun}, {Bridge},
  {Charmandaris}, {Jensen}, {Kewley}, {Lord}, {Madore}, {Marshall},
  {Melbourne}, {Rich}, {Satyapal}, {Schulz}, {Spoon}, {Sturm}, {U}, {Veilleux},
  \& {Xu}}]{Armus09}
{Armus}, L., {Mazzarella}, J.~M., {Evans}, A.~S., {et~al.} 2009, \pasp, 121,
  559

\bibitem[{{Beifiori} {et~al.}(2012){Beifiori}, {Courteau}, {Corsini}, \&
  {Zhu}}]{Beifiori12}
{Beifiori}, A., {Courteau}, S., {Corsini}, E.~M., \& {Zhu}, Y. 2012, \mnras,
  419, 2497

\bibitem[{{Bennert} {et~al.}(2011){Bennert}, {Auger}, {Treu}, {Woo}, \&
  {Malkan}}]{Bennert11}
{Bennert}, V.~N., {Auger}, M.~W., {Treu}, T., {Woo}, J.-H., \& {Malkan}, M.~A.
  2011, \apj, 726, 59

\bibitem[{{Bondi}(1952)}]{Bondi52}
{Bondi}, H. 1952, \mnras, 112, 195

\bibitem[{{Bondi} \& {Hoyle}(1944)}]{Bondi44}
{Bondi}, H., \& {Hoyle}, F. 1944, \mnras, 104, 273

\bibitem[{{Borys} {et~al.}(2005){Borys}, {Smail}, {Chapman}, {Blain},
  {Alexander}, \& {Ivison}}]{Borys05}
{Borys}, C., {Smail}, I., {Chapman}, S.~C., {et~al.} 2005, \apj, 635, 853

\bibitem[{{Cappellari}(2008)}]{JAM}
{Cappellari}, M. 2008, \mnras, 390, 71

\bibitem[{{Cappellari} \& {Copin}(2003)}]{voronoi}
{Cappellari}, M., \& {Copin}, Y. 2003, \mnras, 342, 345

\bibitem[{{Cappellari} \& {Emsellem}(2004)}]{pPXF}
{Cappellari}, M., \& {Emsellem}, E. 2004, \pasp, 116, 138

\bibitem[{{Cen}(2012)}]{Cen12}
{Cen}, R. 2012, \apj, 755, 28

\bibitem[{{Cisternas} {et~al.}(2011{\natexlab{a}}){Cisternas}, {Jahnke},
  {Bongiorno}, {Inskip}, {Impey}, {Koekemoer}, {Merloni}, {Salvato}, \&
  {Trump}}]{Cisternas11}
{Cisternas}, M., {Jahnke}, K., {Bongiorno}, A., {et~al.} 2011{\natexlab{a}},
  \apjl, 741, L11

\bibitem[{{Cisternas} {et~al.}(2011{\natexlab{b}}){Cisternas}, {Jahnke},
  {Inskip}, {Kartaltepe}, {Koekemoer}, {Lisker}, {Robaina}, {Scodeggio},
  {Sheth}, {Trump}, {Andrae}, {Miyaji}, {Lusso}, {Brusa}, {Capak},
  {Cappelluti}, {Civano}, {Ilbert}, {Impey}, {Leauthaud}, {Lilly}, {Salvato},
  {Scoville}, \& {Taniguchi}}]{Cisternas11_merger}
{Cisternas}, M., {Jahnke}, K., {Inskip}, K.~J., {et~al.} 2011{\natexlab{b}},
  \apj, 726, 57

\bibitem[{{Dasyra} {et~al.}(2006){Dasyra}, {Tacconi}, {Davies}, {Naab},
  {Genzel}, {Lutz}, {Sturm}, {Baker}, {Veilleux}, {Sanders}, \&
  {Burkert}}]{Dasyra06}
{Dasyra}, K.~M., {Tacconi}, L.~J., {Davies}, R.~I., {et~al.} 2006, \apj, 651,
  835

\bibitem[{{Davies}(2007)}]{Davies07}
{Davies}, R.~I. 2007, \mnras, 375, 1099

\bibitem[{{Denney} {et~al.}(2009){Denney}, {Watson}, {Peterson}, {Pogge},
  {Atlee}, {Bentz}, {Bird}, {Brokofsky}, {Comins}, {Dietrich}, {Doroshenko},
  {Eastman}, {Efimov}, {Gaskell}, {Hedrick}, {Klimanov}, {Klimek}, {Kruse},
  {Lamb}, {Leighly}, {Minezaki}, {Nazarov}, {Petersen}, {Peterson},
  {Poindexter}, {Schlesinger}, {Sakata}, {Sergeev}, {Tobin}, {Unterborn},
  {Vestergaard}, {Watkins}, \& {Yoshii}}]{Reverb}
{Denney}, K.~D., {Watson}, L.~C., {Peterson}, B.~M., {et~al.} 2009, \apj, 702,
  1353

\bibitem[{{Ellison} {et~al.}(2013){Ellison}, {Mendel}, {Scudder}, {Patton}, \&
  {Palmer}}]{Ellison13}
{Ellison}, S.~L., {Mendel}, J.~T., {Scudder}, J.~M., {Patton}, D.~R., \&
  {Palmer}, M.~J.~D. 2013, \mnras, 430, 3128

\bibitem[{{Ellison} {et~al.}(2011){Ellison}, {Patton}, {Mendel}, \&
  {Scudder}}]{Ellison11}
{Ellison}, S.~L., {Patton}, D.~R., {Mendel}, J.~T., \& {Scudder}, J.~M. 2011,
  \mnras, 418, 2043

\bibitem[{{Ferrarese} \& {Merritt}(2000)}]{FerrareseMerritt00}
{Ferrarese}, L., \& {Merritt}, D. 2000, \apjl, 539, L9

\bibitem[{{Gebhardt} {et~al.}(2000){Gebhardt}, {Bender}, {Bower}, {Dressler},
  {Faber}, {Filippenko}, {Green}, {Grillmair}, {Ho}, {Kormendy}, {Lauer},
  {Magorrian}, {Pinkney}, {Richstone}, \& {Tremaine}}]{Gebhardt00}
{Gebhardt}, K., {Bender}, R., {Bower}, G., {et~al.} 2000, \apjl, 539, L13

\bibitem[{{Gleason}(1988)}]{Gleason88}
{Gleason}, J.~R. 1988, American Statistician, 42, 263

\bibitem[Graham 
\& Scott(2013)]{GrahamScott13} Graham, A.~W., \& Scott, N.\ 2013, \apj, 764, 151 

\bibitem[{{G{\"u}ltekin} {et~al.}(2009){G{\"u}ltekin}, {Richstone}, {Gebhardt},
  {Lauer}, {Pinkney}, {Aller}, {Bender}, {Dressler}, {Faber}, {Filippenko},
  {Green}, {Ho}, {Kormendy}, \& {Siopis}}]{Gult_nuker}
{G{\"u}ltekin}, K., {Richstone}, D.~O., {Gebhardt}, K., {et~al.} 2009, \apj,
  695, 1577

\bibitem[{{Haan} {et~al.}(2011){Haan}, {Surace}, {Armus}, {Evans}, {Howell},
  {Mazzarella}, {Kim}, {Vavilkin}, {Inami}, {Sanders}, {Petric}, {Bridge},
  {Melbourne}, {Charmandaris}, {Diaz-Santos}, {Murphy}, {U}, {Stierwalt}, \&
  {Marshall}}]{Haan11}
{Haan}, S., {Surace}, J.~A., {Armus}, L., {et~al.} 2011, \aj, 141, 100

\bibitem[{{H{\"a}ussler} {et~al.}(2007){H{\"a}ussler}, {McIntosh}, {Barden},
  {Bell}, {Rix}, {Borch}, {Beckwith}, {Caldwell}, {Heymans}, {Jahnke}, {Jogee},
  {Koposov}, {Meisenheimer}, {S{\'a}nchez}, {Somerville}, {Wisotzki}, \&
  {Wolf}}]{Haussler07}
{H{\"a}ussler}, B., {McIntosh}, D.~H., {Barden}, M., {et~al.} 2007, \apjs, 172,
  615

\bibitem[{{Hinshaw} {et~al.}(2009){Hinshaw}, {Weiland}, {Hill}, {Odegard},
  {Larson}, {Bennett}, {Dunkley}, {Gold}, {Greason}, {Jarosik}, {Komatsu},
  {Nolta}, {Page}, {Spergel}, {Wollack}, {Halpern}, {Kogut}, {Limon}, {Meyer},
  {Tucker}, \& {Wright}}]{Hinshaw09}
{Hinshaw}, G., {Weiland}, J.~L., {Hill}, R.~S., {et~al.} 2009, \apjs, 180, 225

\bibitem[{{Hopkins}(2012)}]{Hopkins12}
{Hopkins}, P.~F. 2012, \mnras, 420, L8

\bibitem[{{Hopkins} \& {Elvis}(2010)}]{Hopkins10}
{Hopkins}, P.~F., \& {Elvis}, M. 2010, \mnras, 401, 7

\bibitem[{{Hopkins} {et~al.}(2005){Hopkins}, {Hernquist}, {Cox}, {Di Matteo},
  {Martini}, {Robertson}, \& {Springel}}]{Hopkins05}
{Hopkins}, P.~F., {Hernquist}, L., {Cox}, T.~J., {et~al.} 2005, \apj, 630, 705

\bibitem[{{Hopkins} {et~al.}(2006){Hopkins}, {Hernquist}, {Cox}, {Di Matteo},
  {Robertson}, \& {Springel}}]{Hopkins06}
---. 2006, \apjs, 163, 1

\bibitem[{{Hopkins} {et~al.}(2007{\natexlab{a}}){Hopkins}, {Hernquist}, {Cox},
  {Robertson}, \& {Krause}}]{Hopkins07a}
{Hopkins}, P.~F., {Hernquist}, L., {Cox}, T.~J., {Robertson}, B., \& {Krause},
  E. 2007{\natexlab{a}}, \apj, 669, 45

\bibitem[{{Hopkins} {et~al.}(2007{\natexlab{b}}){Hopkins}, {Hernquist}, {Cox},
  {Robertson}, \& {Krause}}]{Hopkins07b}
---. 2007{\natexlab{b}}, \apj, 669, 67

\bibitem[{{Hopkins} {et~al.}(2009){Hopkins}, {Murray}, \&
  {Thompson}}]{Hopkins09}
{Hopkins}, P.~F., {Murray}, N., \& {Thompson}, T.~A. 2009, \mnras, 398, 303

\bibitem[{{Hopkins} \& {Quataert}(2010{\natexlab{a}})}]{Hopkins10_BHaccretion}
{Hopkins}, P.~F., \& {Quataert}, E. 2010{\natexlab{a}}, \mnras, 407, 1529

\bibitem[{{Hopkins} \&
  {Quataert}(2010{\natexlab{b}})}]{Hopkins10_andromedadisk}
---. 2010{\natexlab{b}}, \mnras, 405, L41

\bibitem[{{Hopkins} \& {Quataert}(2011)}]{Hopkins11}
---. 2011, \mnras, 415, 1027

\bibitem[{{Howell} {et~al.}(2010){Howell}, {Armus}, {Mazzarella}, {Evans},
  {Surace}, {Sanders}, {Petric}, {Appleton}, {Bothun}, {Bridge}, {Chan},
  {Charmandaris}, {Frayer}, {Haan}, {Inami}, {Kim}, {Lord}, {Madore},
  {Melbourne}, {Schulz}, {U}, {Vavilkin}, {Veilleux}, \& {Xu}}]{Howell10}
{Howell}, J.~H., {Armus}, L., {Mazzarella}, J.~M., {et~al.} 2010, \apj, 715,
  572

\bibitem[{{Ishida}(2004)}]{IshidaPhD}
{Ishida}, C.~M. 2004, PhD thesis, Institute for Astronomy, University of
  Hawaii, 2680 Woodlawn Dr.~Honolulu, HI 96822

\bibitem[{{Jahnke} {et~al.}(2009){Jahnke}, {Bongiorno}, {Brusa}, {Capak},
  {Cappelluti}, {Cisternas}, {Civano}, {Colbert}, {Comastri}, {Elvis},
  {Hasinger}, {Ilbert}, {Impey}, {Inskip}, {Koekemoer}, {Lilly}, {Maier},
  {Merloni}, {Riechers}, {Salvato}, {Schinnerer}, {Scoville}, {Silverman},
  {Taniguchi}, {Trump}, \& {Yan}}]{Jahnke09}
{Jahnke}, K., {Bongiorno}, A., {Brusa}, M., {et~al.} 2009, \apjl, 706, L215

\bibitem[{{Kauffmann} \& {Heckman}(2009)}]{KauffmannHeckman09}
{Kauffmann}, G., \& {Heckman}, T.~M. 2009, \mnras, 397, 135

\bibitem[{{Kim} {et~al.}(2013){Kim}, {Evans}, {Vavilkin}, {Armus},
  {Mazzarella}, {Sheth}, {Surace}, {Haan}, {Howell}, {D{\'{\i}}az-Santos},
  {Petric}, {Iwasawa}, {Privon}, \& {Sanders}}]{Kim13}
{Kim}, D.-C., {Evans}, A.~S., {Vavilkin}, T., {et~al.} 2013, \apj, 768, 102

\bibitem[{{King}(2008)}]{King08}
{King}, A. 2008, \nar, 52, 253

\bibitem[{{Kl{\"o}ckner} \& {Baan}(2004)}]{Klockner04}
{Kl{\"o}ckner}, H.-R., \& {Baan}, W.~A. 2004, \aap, 419, 887

\bibitem[{{Kocevski} {et~al.}(2012){Kocevski}, {Faber}, {Mozena}, {Koekemoer},
  {Nandra}, {Rangel}, {Laird}, {Brusa}, {Wuyts}, {Trump}, {Koo}, {Somerville},
  {Bell}, {Lotz}, {Alexander}, {Bournaud}, {Conselice}, {Dahlen}, {Dekel},
  {Donley}, {Dunlop}, {Finoguenov}, {Georgakakis}, {Giavalisco}, {Guo},
  {Grogin}, {Hathi}, {Juneau}, {Kartaltepe}, {Lucas}, {McGrath}, {McIntosh},
  {Mobasher}, {Robaina}, {Rosario}, {Straughn}, {van der Wel}, \&
  {Villforth}}]{Kocevski12}
{Kocevski}, D.~D., {Faber}, S.~M., {Mozena}, M., {et~al.} 2012, \apj, 744, 148

\bibitem[{{Kormendy} \& {Gebhardt}(2001)}]{KormendyGebhardt01}
{Kormendy}, J., \& {Gebhardt}, K. 2001, in American Institute of Physics
  Conference Series, Vol. 586, 20th Texas Symposium on relativistic
  astrophysics, ed. J.~C. {Wheeler} \& H.~{Martel}, 363--381

\bibitem[{{Kormendy} \& {Ho}(2013)}]{KormendyHo13}
{Kormendy}, J., \& {Ho}, L.~C. 2013, \araa, 51, 511

\bibitem[{{Kormendy} \& {Richstone}(1995)}]{KormendyRichstone95}
{Kormendy}, J., \& {Richstone}, D. 1995, \araa, 33, 581

\bibitem[{{Koss} {et~al.}(2013){Koss}, {Mushotzky}, {Baumgartner}, {Veilleux},
  {Tueller}, {Markwardt}, \& {Casey}}]{Koss13}
{Koss}, M., {Mushotzky}, R., {Baumgartner}, W., {et~al.} 2013, \apjl, 765, L26

\bibitem[{{Koss} {et~al.}(2010){Koss}, {Mushotzky}, {Veilleux}, \&
  {Winter}}]{Koss10}
{Koss}, M., {Mushotzky}, R., {Veilleux}, S., \& {Winter}, L. 2010, \apjl, 716,
  L125

\bibitem[{{Larkin} {et~al.}(2006){Larkin}, {Barczys}, {Krabbe}, {Adkins},
  {Aliado}, {Amico}, {Brims}, {Campbell}, {Canfield}, {Gasaway}, {Honey},
  {Iserlohe}, {Johnson}, {Kress}, {LaFreniere}, {Lyke}, {Magnone}, {Magnone},
  {McElwain}, {Moon}, {Quirrenbach}, {Skulason}, {Song}, {Spencer}, {Weiss}, \&
  {Wright}}]{Larkin06}
{Larkin}, J., {Barczys}, M., {Krabbe}, A., {et~al.} 2006, in Society of
  Photo-Optical Instrumentation Engineers (SPIE) Conference Series, Vol. 6269,
  Society of Photo-Optical Instrumentation Engineers (SPIE) Conference Series

\bibitem[{{Magorrian} {et~al.}(1998){Magorrian}, {Tremaine}, {Richstone},
  {Bender}, {Bower}, {Dressler}, {Faber}, {Gebhardt}, {Green}, {Grillmair},
  {Kormendy}, \& {Lauer}}]{Magorrian98}
{Magorrian}, J., {Tremaine}, S., {Richstone}, D., {et~al.} 1998, \aj, 115, 2285

\bibitem[{{Marconi} \& {Hunt}(2003)}]{MarconiHunt03}
{Marconi}, A., \& {Hunt}, L.~K. 2003, \apjl, 589, L21

\bibitem[{{McConnell} \& {Ma}(2013)}]{McConnell13}
{McConnell}, N.~J., \& {Ma}, C.-P. 2013, \apj, 764, 184

\bibitem[{{Medling} {et~al.}(2011){Medling}, {Ammons}, {Max}, {Davies},
  {Engel}, \& {Canalizo}}]{medling11}
{Medling}, A.~M., {Ammons}, S.~M., {Max}, C.~E., {et~al.} 2011, \apj, 743, 32

\bibitem[{{Medling} {et~al.}(2014){Medling}, {U}, {Guedes}, {Max}, {Mayer},
  {Armus}, {Holden}, {Ro{\v s}kar}, \& {Sanders}}]{nucleardisks}
{Medling}, A.~M., {U}, V., {Guedes}, J., {et~al.} 2014, \apj, 784, 70

\bibitem[{{Murray}(2009)}]{Murray09}
{Murray}, N. 2009, \apj, 691, 946

\bibitem[{{Peng} {et~al.}(2002){Peng}, {Ho}, {Impey}, \& {Rix}}]{Peng02}
{Peng}, C.~Y., {Ho}, L.~C., {Impey}, C.~D., \& {Rix}, H.-W. 2002, \aj, 124, 266

\bibitem[{{Peng} {et~al.}(2010){Peng}, {Ho}, {Impey}, \& {Rix}}]{Peng10}
---. 2010, \aj, 139, 2097

\bibitem[{{Power} {et~al.}(2011){Power}, {Nayakshin}, \& {King}}]{Power11}
{Power}, C., {Nayakshin}, S., \& {King}, A. 2011, \mnras, 412, 269

\bibitem[{{Privon} {et~al.}(2013){Privon}, {Barnes}, {Evans}, {Hibbard}, {Yun},
  {Mazzarella}, {Armus}, \& {Surace}}]{Privon13}
{Privon}, G.~C., {Barnes}, J.~E., {Evans}, A.~S., {et~al.} 2013, \apj, 771, 120

\bibitem[{{Rothberg} \& {Joseph}(2006)}]{Rothberg06}
{Rothberg}, B., \& {Joseph}, R.~D. 2006, \aj, 131, 185

\bibitem[{{Rupke} \& {Veilleux}(2011)}]{Rupke11}
{Rupke}, D.~S.~N., \& {Veilleux}, S. 2011, \apjl, 729, L27

\bibitem[{{Sanders} \& {Mirabel}(1996)}]{Sanders96}
{Sanders}, D.~B., \& {Mirabel}, I.~F. 1996, \araa, 34, 749

\bibitem[{{Sanders} {et~al.}(1988){Sanders}, {Soifer}, {Elias}, {Madore},
  {Matthews}, {Neugebauer}, \& {Scoville}}]{Sanders88}
{Sanders}, D.~B., {Soifer}, B.~T., {Elias}, J.~H., {et~al.} 1988, \apj, 325, 74

\bibitem[{{Scoville} {et~al.}(2000){Scoville}, {Evans}, {Thompson}, {Rieke},
  {Hines}, {Low}, {Dinshaw}, {Surace}, \& {Armus}}]{Scoville00}
{Scoville}, N.~Z., {Evans}, A.~S., {Thompson}, R., {et~al.} 2000, \aj, 119, 991

\bibitem[{{Shier} {et~al.}(1996){Shier}, {Rieke}, \& {Rieke}}]{Shier96}
{Shier}, L.~M., {Rieke}, M.~J., \& {Rieke}, G.~H. 1996, \apj, 470, 222

\bibitem[{{Siopis} {et~al.}(2009){Siopis}, {Gebhardt}, {Lauer}, {Kormendy},
  {Pinkney}, {Richstone}, {Faber}, {Tremaine}, {Aller}, {Bender}, {Bower},
  {Dressler}, {Filippenko}, {Green}, {Ho}, \& {Magorrian}}]{Siopis}
{Siopis}, C., {Gebhardt}, K., {Lauer}, T.~R., {et~al.} 2009, \apj, 693, 946

\bibitem[{{Spoon} {et~al.}(2013){Spoon}, {Farrah}, {Lebouteiller},
  {Gonz{\'a}lez-Alfonso}, {Bernard-Salas}, {Urrutia}, {Rigopoulou},
  {Westmoquette}, {Smith}, {Afonso}, {Pearson}, {Cormier}, {Efstathiou},
  {Borys}, {Verma}, {Etxaluze}, \& {Clements}}]{Spoon13}
{Spoon}, H.~W.~W., {Farrah}, D., {Lebouteiller}, V., {et~al.} 2013, \apj, 775,
  127

\bibitem[{{Springel} {et~al.}(2005{\natexlab{a}}){Springel}, {Di Matteo}, \&
  {Hernquist}}]{Springel05}
{Springel}, V., {Di Matteo}, T., \& {Hernquist}, L. 2005{\natexlab{a}}, \apjl,
  620, L79

\bibitem[{{Springel} {et~al.}(2005{\natexlab{b}}){Springel}, {Di Matteo}, \&
  {Hernquist}}]{Springel05_feedback}
---. 2005{\natexlab{b}}, \mnras, 361, 776

\bibitem[{{Stickley} \& {Canalizo}(2014)}]{Stickley14}
{Stickley}, N.~R., \& {Canalizo}, G. 2014, \apj, 786, 12

\bibitem[{{Sturm} {et~al.}(2011){Sturm}, {Gonz{\'a}lez-Alfonso}, {Veilleux},
  {Fischer}, {Graci{\'a}-Carpio}, {Hailey-Dunsheath}, {Contursi}, {Poglitsch},
  {Sternberg}, {Davies}, {Genzel}, {Lutz}, {Tacconi}, {Verma}, {Maiolino}, \&
  {de Jong}}]{Sturm11}
{Sturm}, E., {Gonz{\'a}lez-Alfonso}, E., {Veilleux}, S., {et~al.} 2011, \apjl,
  733, L16

\bibitem[{{Tecza} {et~al.}(2000){Tecza}, {Genzel}, {Tacconi}, {Anders},
  {Tacconi-Garman}, \& {Thatte}}]{Tecza}
{Tecza}, M., {Genzel}, R., {Tacconi}, L.~J., {et~al.} 2000, \apj, 537, 178

\bibitem[{{Tremaine} {et~al.}(2002){Tremaine}, {Gebhardt}, {Bender}, {Bower},
  {Dressler}, {Faber}, {Filippenko}, {Green}, {Grillmair}, {Ho}, {Kormendy},
  {Lauer}, {Magorrian}, {Pinkney}, \& {Richstone}}]{Tremaine02}
{Tremaine}, S., {Gebhardt}, K., {Bender}, R., {et~al.} 2002, \apj, 574, 740

\bibitem[{{Tremonti} {et~al.}(2004){Tremonti}, {Heckman}, {Kauffmann},
  {Brinchmann}, {Charlot}, {White}, {Seibert}, {Peng}, {Schlegel}, {Uomoto},
  {Fukugita}, \& {Brinkmann}}]{Tremonti04}
{Tremonti}, C.~A., {Heckman}, T.~M., {Kauffmann}, G., {et~al.} 2004, \apj, 613,
  898

\bibitem[{{U} {et~al.}(2012){U}, {Sanders}, {Mazzarella}, {Evans}, {Howell},
  {Surace}, {Armus}, {Iwasawa}, {Kim}, {Casey}, {Vavilkin}, {Dufault},
  {Larson}, {Barnes}, {Chan}, {Frayer}, {Haan}, {Inami}, {Ishida},
  {Kartaltepe}, {Melbourne}, \& {Petric}}]{U_SED}
{U}, V., {Sanders}, D.~B., {Mazzarella}, J.~M., {et~al.} 2012, \apjs, 203, 9

\bibitem[{{U} {et~al.}(2013){U}, {Medling}, {Sanders}, {Max}, {Armus},
  {Iwasawa}, {Evans}, {Kewley}, \& {Fazio}}]{mrk273}
{U}, V., {Medling}, A.~M., {Sanders}, D., {et~al.} 2013, \apj, 775, 115

\bibitem[{{van Dam} {et~al.}(2004){van Dam}, {Le Mignant}, \&
  {Macintosh}}]{vanDam04}
{van Dam}, M.~A., {Le Mignant}, D., \& {Macintosh}, B.~A. 2004, \ao, 43, 5458

\bibitem[{{van Dam} {et~al.}(2006){van Dam}, {Bouchez}, {Le Mignant},
  {Johansson}, {Wizinowich}, {Campbell}, {Chin}, {Hartman}, {Lafon}, {Stomski},
  \& {Summers}}]{vanDam06}
{van Dam}, M.~A., {Bouchez}, A.~H., {Le Mignant}, D., {et~al.} 2006, \pasp,
  118, 310

\bibitem[{{Veilleux} {et~al.}(2002){Veilleux}, {Kim}, \&
  {Sanders}}]{Veilleux02}
{Veilleux}, S., {Kim}, D.-C., \& {Sanders}, D.~B. 2002, \apjs, 143, 315

\bibitem[{{Veilleux} {et~al.}(2013){Veilleux}, {Mel{\'e}ndez}, {Sturm},
  {Gracia-Carpio}, {Fischer}, {Gonz{\'a}lez-Alfonso}, {Contursi}, {Lutz},
  {Poglitsch}, {Davies}, {Genzel}, {Tacconi}, {de Jong}, {Sternberg}, {Netzer},
  {Hailey-Dunsheath}, {Verma}, {Rupke}, {Maiolino}, {Teng}, \&
  {Polisensky}}]{Veilleux13}
{Veilleux}, S., {Mel{\'e}ndez}, M., {Sturm}, E., {et~al.} 2013, \apj, 776, 27

\bibitem[{{Wilson} {et~al.}(2008){Wilson}, {Petitpas}, {Iono}, {Baker}, {Peck},
  {Krips}, {Warren}, {Golding}, {Atkinson}, {Armus}, {Cox}, {Ho}, {Juvela},
  {Matsushita}, {Mihos}, {Pihlstrom}, \& {Yun}}]{Wilson08}
{Wilson}, C.~D., {Petitpas}, G.~R., {Iono}, D., {et~al.} 2008, \apjs, 178, 189

\bibitem[{{Winge} {et~al.}(2009){Winge}, {Riffel}, \&
  {Storchi-Bergmann}}]{GNIRS}
{Winge}, C., {Riffel}, R.~A., \& {Storchi-Bergmann}, T. 2009, \apjs, 185, 186

\bibitem[{{Wizinowich} {et~al.}(2000){Wizinowich}, {Acton}, {Lai}, {Gathright},
  {Lupton}, \& {Stomski}}]{Wiz00}
{Wizinowich}, P.~L., {Acton}, D.~S., {Lai}, O., {et~al.} 2000, in Society of
  Photo-Optical Instrumentation Engineers (SPIE) Conference Series, Vol. 4007,
  Society of Photo-Optical Instrumentation Engineers (SPIE) Conference Series,
  ed. P.~L. {Wizinowich}, 2--13

\bibitem[{{Wizinowich} {et~al.}(2006){Wizinowich}, {Le Mignant}, {Bouchez},
  {Campbell}, {Chin}, {Contos}, {van Dam}, {Hartman}, {Johansson}, {Lafon},
  {Lewis}, {Stomski}, {Summers}, {Brown}, {Danforth}, {Max}, \&
  {Pennington}}]{Wiz06}
{Wizinowich}, P.~L., {Le Mignant}, D., {Bouchez}, A.~H., {et~al.} 2006, \pasp,
  118, 297

\bibitem[{{Wurster} \& {Thacker}(2013{\natexlab{a}})}]{Wurster13_comparison}
{Wurster}, J., \& {Thacker}, R.~J. 2013{\natexlab{a}}, \mnras, 431, 2513

\bibitem[{{Wurster} \& {Thacker}(2013{\natexlab{b}})}]{Wurster13}
---. 2013{\natexlab{b}}, \mnras, 431, 539

\bibitem[{{Zhang} {et~al.}(2012){Zhang}, {Lu}, \& {Yu}}]{Zhang12}
{Zhang}, X., {Lu}, Y., \& {Yu}, Q. 2012, \apj, 761, 5

\end{thebibliography}




\end{document}